\documentclass[nonacm,sigplan]{acmart}
\AtBeginDocument{%
  }

\usepackage{verbatim}
\usepackage{multirow}
\usepackage{array}
\usepackage{longtable}
\usepackage{booktabs}
\usepackage{hyperref}
\usepackage{caption}
\usepackage{tabularx}
\usepackage{graphicx}
\usepackage{multicol}
\usepackage{geometry}
\usepackage{url}
\usepackage{lscape}
\usepackage{xcolor}
\usepackage{float}

\makeatletter
\def\@copyrightspace{\relax}
\makeatother

\setcopyright{none} 
\renewcommand\footnotetextcopyrightpermission[1]{} % removes footnote with conference information in first column

\begin{document}

\title{A PRISMA-Driven Bibliometric Analysis of the Scientific Literature on Assurance Case Patterns}


\author{Oluwafemi Odu}
\affiliation{%
  \institution{York University}
  \city{Toronto}
  \country{Canada}
}
\email{olufemi2@yorku.ca}

\author{Alvine Boaye Belle}
\affiliation{%
  \institution{York University}
  \city{Toronto}
  \country{Canada}
}
\email{alvine.belle@lassonde.yorku.ca}

\author{Song Wang}
\affiliation{%
  \institution{York University}
  \city{Toronto}
  \country{Canada}
}
\email{wangsong@yorku.ca}

\author{Kimya Khakzad Shahandashti}
\affiliation{%
  \institution{York University}
  \city{Toronto}
  \country{Canada}
}
\email{kimya@yorku.ca}

\renewcommand{\shortauthors}{odu et al.}


\begin{abstract}

Justifying the correct implementation of the non-functional requirements (e.g., safety, security) of mission-critical systems is crucial to prevent system failure. The latter could have severe consequences such as the death of people and financial losses. Assurance cases can be used to prevent system failure. They are structured arguments that allow arguing and relaying various safety-critical systems’ requirements extensively, as well as checking the compliance of such systems with industrial standards to support their certification. Still, the creation of assurance cases is usually manual, error-prone, and time-consuming. Besides, it may involve numerous alterations as the system evolves. To overcome the bottlenecks in creating assurance cases, existing approaches usually promote the reuse of common structured evidence-based arguments (i.e. patterns) to aid the creation of assurance cases. To gain insights into the advancements of the research on assurance case patterns, we relied on SEGRESS to conduct a bibliometric analysis of 92 primary studies published within the past two decades. This allows capturing the evolutionary trends and patterns characterizing the research in that field. Our findings notably indicate the emergence of new assurance case patterns to support the assurance of ML-enabled systems that are characterized by their evolving requirements (e.g., cybersecurity and ethics). 

\end{abstract}




\begin{comment}

\begin{CCSXML}
<ccs2012>
 <concept>
  <concept_id>00000000.0000000.0000000</concept_id>
  <concept_desc>Do Not Use This Code, Generate the Correct Terms for Your Paper</concept_desc>
  <concept_significance>500</concept_significance>
 </concept>
 <concept>
  <concept_id>00000000.00000000.00000000</concept_id>
  <concept_desc>Do Not Use This Code, Generate the Correct Terms for Your Paper</concept_desc>
  <concept_significance>300</concept_significance>
 </concept>
 <concept>
  <concept_id>00000000.00000000.00000000</concept_id>
  <concept_desc>Do Not Use This Code, Generate the Correct Terms for Your Paper</concept_desc>
  <concept_significance>100</concept_significance>
 </concept>
 <concept>
  <concept_id>00000000.00000000.00000000</concept_id>
  <concept_desc>Do Not Use This Code, Generate the Correct Terms for Your Paper</concept_desc>
  <concept_significance>100</concept_significance>
 </concept>
</ccs2012>
\end{CCSXML}

\ccsdesc[500]{Do Not Use This Code~Generate the Correct Terms for Your Paper}
\ccsdesc[300]{Do Not Use This Code~Generate the Correct Terms for Your Paper}
\ccsdesc{Do Not Use This Code~Generate the Correct Terms for Your Paper}
\ccsdesc[100]{Do Not Use This Code~Generate the Correct Terms for Your Paper}
    
\end{comment}



\maketitle

\section{INTRODUCTION}

Mission-critical systems are increasingly designed to be interoperable and interconnected. Their operational contexts usually changes at runtime \cite{b8,b35,b73}. Hence, these systems usually operate under unpredictable conditions throughout their life cycle. That unpredictability is caused by the limited control they have over their environment, the dynamic emergence of new types of objects and events in the world, as well as the increasing growth of the number of composite configurations of such elements. Justifying and providing confidence in the essential properties (e.g., security, safety) of these mission-critical systems is therefore crucial to prevent system failure. The latter could result in severe injuries, death of people, and property destruction, just to name a few~\cite{b7}.

Assurance cases are structured arguments that allow justifying that the properties of a system have been correctly implemented ~\cite{b1}. 
Assurance cases are utilized across various domains (e.g., medicine \cite{b2,b3,b4}, automotive \cite{b5,b6,b7}, aerospace) to demonstrate the reliability of mission-critical systems and support their certification in compliance with industry standards (e.g., ISO 26262 \cite{b131}, DO-178C \cite{b132}). In safety-critical domains, it is essential to ensure the system’s safety when there is a change in its operational context~\cite{b8}. Navigating these complexities and ensuring system reassurance with each change in the operating context can be time-consuming, tedious, and expensive \cite{b8, b72}. 

The introduction of assurance case patterns (ACPs) in \cite{b9}, as a template formed from previous successful assurance cases, has facilitated the reuse of these templates for creating new assurance cases. Assurance case patterns are instantiated with system-specific information during the creation of an assurance case for a given system. 
Despite the wide use of assurance case patterns over the years to support the automatic generation of assurance cases, there has been -- to the best of our knowledge -- no review of assurance case patterns from a bibliometric analysis perspective. This makes it challenging to assess the evolution of the associated line of research over the years.  

Bibliometric analysis is a robust method for analyzing large volumes of scientific data to uncover the evolutionary trends, collaboration patterns, and research structure of a research field~\cite{b14}. To gain insights into the advancements within the field of assurance case patterns, we relied on SEGRESS --a PRISMA 2020 \cite{b43} adaptation -- and common bibliometric analysis techniques \cite{b14} to conduct a bibliometric analysis of \textbf{92} primary studies spanning two decades. We utilized the snowballing technique \cite{b60} and database-driven search to identify these primary studies across five scientific databases.  We analyzed and visualized the annual distribution of publications, publication venues, keyword networks, collaboration maps, and other relevant information extracted from the literature using well-established bibliometric tools (e.g., VosViewer \cite{b15}, Tableau \cite{b16}, Google Charts \cite{b48}), and Microsoft Excel.

In summary, this paper makes the following contributions:
\begin{itemize}
    \item We provide a brief background and survey of related work on assurance case patterns.
    \item We provide an overview of the evolutionary trends, significance, collaboration patterns, and research structure in the field of assurance case patterns. 
    \item We make recommendations based on our results and discuss potential future directions in the field of assurance case patterns. 
\end{itemize}

The rest of the paper is organized as follows. Section~\ref{sec:2} discusses the rationale for our work. Section \ref{sec:3} presents some background concepts in assurance case patterns. Section \ref{sec:4} describes the methodology we used to perform our bibliometric analysis. Section \ref{sec:5} reports the results of our study. Section \ref{sec:6} identifies the threats to validity associated with our study, while section \ref{sec:7} concludes the paper.

\section{RATIONALE FOR CONDUCTING THE BIBLIOMETRIC ANALYSIS}
\label{sec:2}

Bibliometric analyses have been conducted over the years across various domains ranging from agriculture \cite{b18}, medicine \cite{b20}, business \cite{b19}, and tourism \cite{b21}, just to name a few. However, in recent years, there has been a growing recognition of the importance of bibliometric analysis in the software engineering domain \cite{b17, wong2021bibliometric}. For instance,  Karanatsiou et al. \cite{b17} published a bibliometric analysis targeting an eight-year period of 2010---2017, and focusing on software engineering scholars and institutions. They published their work in the Journal of Systems and Software in 2019. Likewise, Wong et al. \cite{wong2021bibliometric} have recently published a follow-up version of the work of Karanatsiou et al. \cite{b17} by carrying out a bibliometric analysis focusing on software engineering themes, scholars, and institutions spanning an eight-year period of 2013---2020.

In the wide and growing field of system assurance, a plethora of secondary studies have also been conducted to understand the current research and growth of the field. 
For instance, Szczygielska and Jarzębowicz \cite{b22} presented an online unified assurance case patterns catalog containing 45 patterns extracted from several sources. Their objective was to aid the quick creation of assurance cases in supporting tools like NOR-STA. Preschern et al. \cite{b23} surveyed the various methods to build or improve a system architecture using safety patterns. They provided an overview of twelve existing pattern-based safety development methods extracted from the literature. They analyzed and compared these methods based on the domain and pattern targeted by these methods, the approach to apply the patterns, the safety standard followed, and the method maturity. Their analysis offers insights into the pros and cons of each pattern-based safety development method. 

 Gleirscher and Kugele \cite{b24} conducted a comprehensive survey examining safety design and argument patterns employed in the construction and assurance of safety-critical systems spanning multiple domains. The study involved addressing twelve survey questions, establishing the correlation between design and argumentation concepts crucial for ensuring system safety. The authors also identified challenges related to the efficient reuse of safety mechanisms within systems and proposed research directions aimed at resolving these challenges.

To verify and assess the benefits relative to the costs associated with the formalization of assurance arguments about a system property, Graydon \cite{b12} conducted a survey encompassing twenty studies focusing on proposed formal assurance arguments. Their findings revealed that a significant portion of these studies primarily speculate on the advantages of formalism without presenting concrete proof to substantiate these presumed advantages. 
Furthermore, driven by the widespread adoption of Model-Based Engineering (MBE) in crafting model-based assurance cases, Yan et al. \cite{b13} surveyed to assess techniques for generating safety cases using MBE. They surveyed and identified a crucial research gap characterized by the lack of automated processing of raw and unstructured instantiable data. To address this gap, the authors suggested the implementation of a System Assurance Case Metamodel (SACM) compliant framework for generating assurance cases.

However, to the best of our knowledge, in the literature of ACPs, there is no paper that has analyzed the state of the art from a bibliometric analysis perspective. This hampers the identification of the emerging trends, research gaps, and potential research directions in the field of assurance case patterns. To tackle that issue, we report a bibliometric analysis that analyzes the entire scientific landscape on ACPs and identifies future research directions in the field of assurance case patterns.

\section{BACKGROUND}
\label{sec:3}

\subsection{What is an assurance case?}
\label{sec:3.1}

\begin{figure*}[hbt!]
    \centering
    \includegraphics[width=1\linewidth]{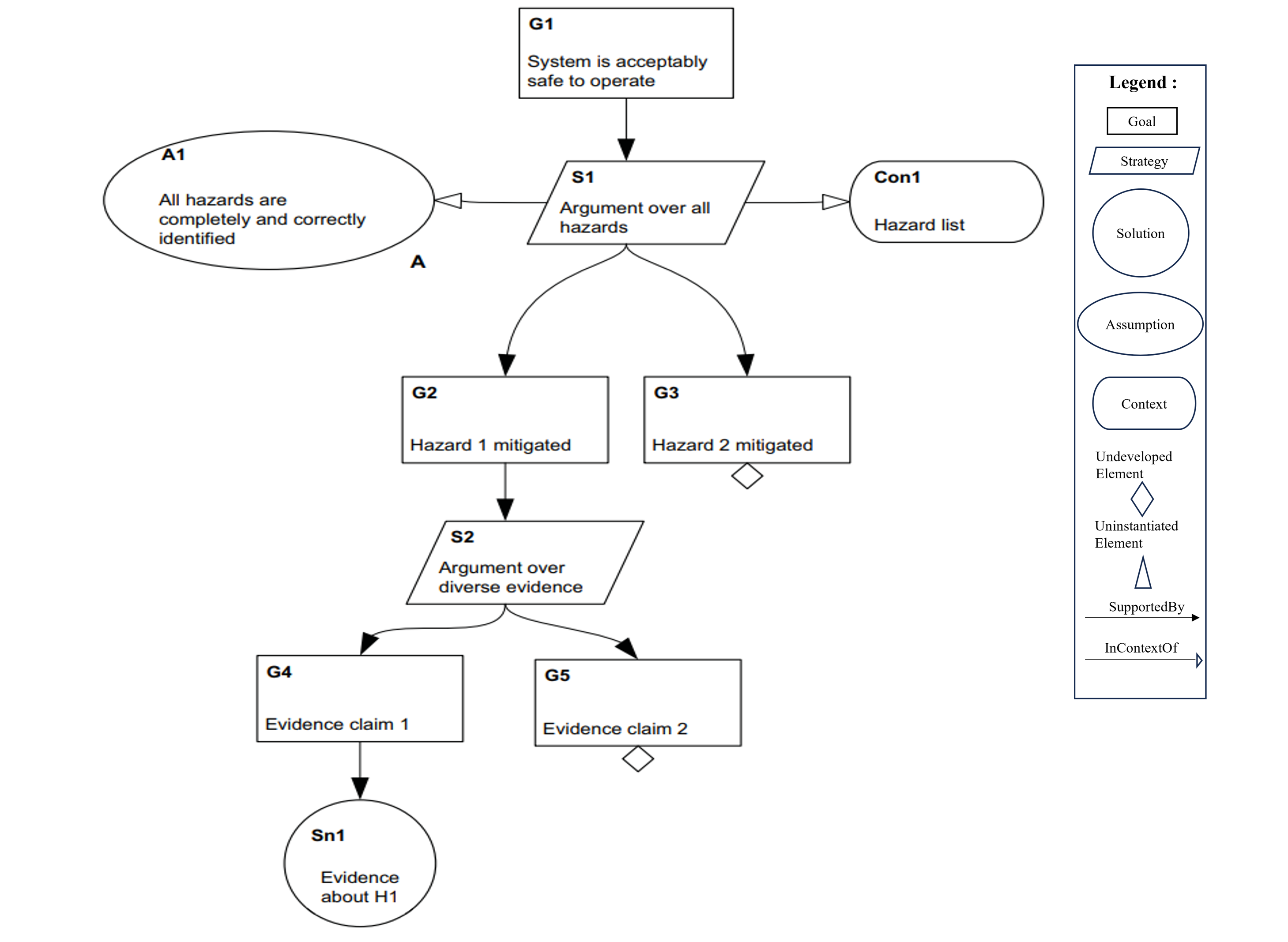}
    \caption{A sample safety case (adapted from \cite{b49}).}
    \label{fig1}
\end{figure*}

An assurance case (AC) is a well-established structured technique used to document a reasoned, auditable argument supporting that a system meets desirable properties \cite{b25}. An assurance case provides a method to justify and access the confidence in a given system property \cite{b1}. 
Assurance case is often used as a broad term. When the focus of assurance is on a specific system property, the name of the property usually precedes to be more specific. Hence, there are several types of assurance cases. These include safety cases \cite{b4, b29}, security cases \cite{b26, b27}, dependability cases \cite{b28}, reliability cases \cite{b30}, and ethics cases \cite{b31, b32}.

Assurance cases are utilized across various domains such as medicine \cite{b2,b3,b4} and automotive \cite{b5,b6,b7} to ensure the reliability of mission-critical systems and support the certification of systems in compliance with industry standards (e.g., ISO 26262, DO-178C). Regulatory agencies such as the Food and Drug Administration (FDA) also recommend the use of assurance cases to support the safety confidence of medical devices during their approval process \cite{b26}. 

In general, the structure of an assurance case often consists of three main parts: (1) a top claim – the top claim is often decomposed into sub-claims. The top claim describes the ultimate or root claim that a system satisfies a given requirement. (2) a set of evidence – which refers to the set of facts, proofs, and system artifacts that provide support to the sub-claims, and (3) a set of structured arguments that links a set of evidence to an associated sub-claim and connects all sub-claims to the top claim of the assurance case~\cite{b130}.

Several notations allow representing assurance cases. These notations include GSN (Goal structuring Notation) \cite{b33}, CAE (Claim-Argument-Evidence) \cite{b1}, SACM (Structured Assurance Case Metamodel) \cite{b34}  and Eliminative Argumentation (EA) \cite{b129}. The Object Management Group (OMG) recently introduced SACM (Structured Assurance Case Metamodel) \cite{b34} to promote interoperability and standardisation \cite{b35}. SACM aligns with existing  assurance case notations (e.g., GSN, CAE). Still, the GSN is the most adopted notation for representing assurance cases \cite{b35}.

GSN supports the representation of an assurance case in a tree-like structure. The GSN community standard \cite{b33} proposes the following six main GSN elements to represent assurance cases: Goal, Strategy, Solution, Context, Assumption, and Justification. 
A goal is depicted as a rectangle and represents the main claim or a sub-claim. A  Strategy is depicted as a parallelogram and describes the inference between a goal and its sub-goals. A Solution is depicted as a circle and represents the evidence supporting an argument or goal. 
Context, Assumption, and Justification are respectively depicted as a rounded rectangle, an oval with the letter \textit{A}, and an oval with the letter \textit{J} respectively. 
The Context, Assumption, and Justification elements specify the context, scope, or rationale of an argument respectively. 

GSN elements can be decorated using the \textit{Undeveloped} and \textit{Uninstantiated} decorators. The \textit{Undeveloped} decorator depicted as a hollow diamond applied to the bottom center of an element, indicates that a GSN element has not been developed \cite{b33}. 
The \textit{Uninstantiated} decorator is depicted with a small triangle applied to the bottom center of an element. This allows indicating that a GSN element is yet to be instantiated (i.e., abstract element in a placeholder needs to be replaced by a concrete instance) \cite{b33}. 

Furthermore, the GSN community standard defines two main relationships between GSN elements which are \textit{SupportedBy}  and \textit{InContextOf}. \textit{SupportedBy} is depicted as a line with a solid arrowhead and represents supporting relationships between GSN elements. \textit{InContextOf} is depicted as a line with a hollow arrowhead and represents a contextual relationship between GSN elements. 
%\newline

Figure \ref{fig1} shows an excerpt assurance case adapted from \cite{b49}. That excerpt is represented in the GSN. 
%The legend of that figure illustrates the main elements of a GSN-based assurance case. 

\subsection{What is an assurance case pattern?}
\begin{figure*}[hbt!]
    \centering
    \includegraphics[width=1\linewidth]{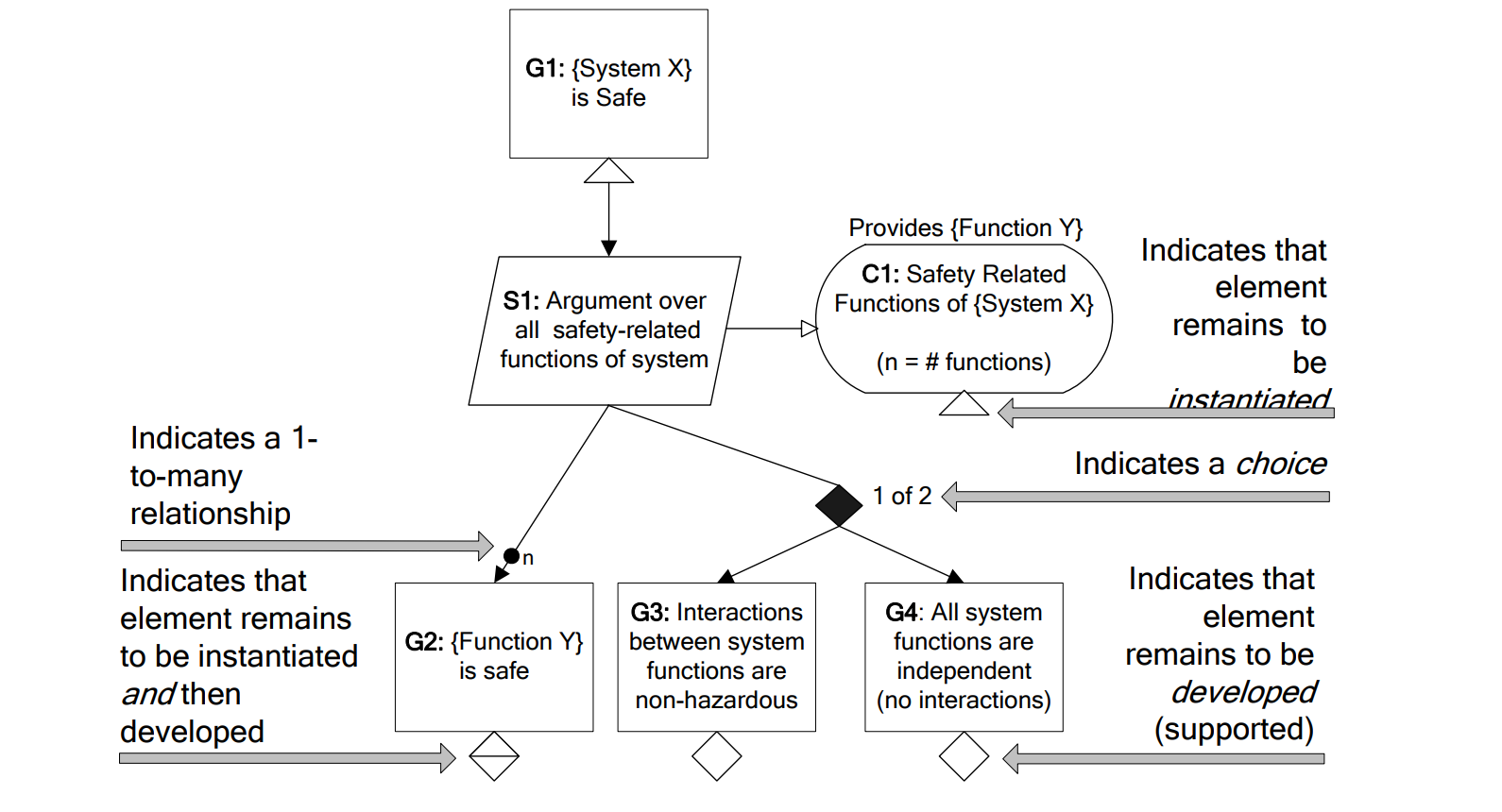}
    \caption{A sample assurance case pattern (adapted from \cite{b61}).}
    \label{fig2}
\end{figure*}

Patterns in simple terms refer to a model or design used to guide the development or construction of an artifact. Assurance case patterns have their origin in design patterns used in software engineering. Kelly introduced assurance case patterns \cite{b9}. An assurance case pattern is a model or template formed from previous successful assurance cases and contains placeholders for system-specific information to aid in re-use and ease the creation of assurance cases~\cite{b8}.  
To create an assurance case from an assurance case pattern, the ACP is instantiated by replacing abstract values in placeholders with concrete values. The abstract values in placeholders represent the generic information present in the main elements of an ACP, while the concrete values represent system-specific information used to replace the generic information in an ACP to create an assurance case.

To support the reuse and automatic creation of assurance cases, the literature has proposed different assurance case patterns \cite{b36,b37,b38} spanning multiple application domains. Additionally, ACPs are also used to mitigate assurance deficits \cite{b38,b40}. Assurance deficits refer to \textit{"any knowledge gap that prohibits total or perfect confidence"} in an assurance case\cite{b49}.

Furthermore, to ensure maximum utilization in the usage of assurance case patterns for creating assurance cases, Denney and Pai \cite{b39} proposed a formal definition of an assurance case pattern. 
Denney and Pai \cite{b39} defined a pattern P, as a tuple \textit{⟨N, l, p, m, c,→ ⟩}, where: \textit{⟨ N,→ ⟩}  is a directed hypergraph in which each hyperedge has a single source and possibly multiple targets; \textit{l} is a labeling function that gives the node type, \textit{t} that gives the node contents, \textit{p} is a parameter label on nodes, \textit{m} gives the multiplicity range on a link between two nodes, and \textit{c} gives the range on the choice attached to a given node.

To allow representing an assurance case pattern, GSN has been extended to support several modeling elements such as multiplicity, optionality, and abstraction \cite{b41}. Hence, in addition to the elements we described in section \ref{sec:3.1}, the GSN extension allows for representing an assurance case pattern by using the following additional symbols: Parameterized expressions, Multiplicity, Optionality, and Choice. Parameterized expressions are abstract expressions inside placeholders that need to be replaced with concrete information \cite{b42}. Multiplicity symbols can be used to describe how many instances of one element type relate to another element. These symbols are generalized n-ary relationships between GSN elements \cite{b33}. Optionality represents optional and alternative relationships between GSN elements which generalizes n-of-m choices between objects \cite{b33}. The choice is depicted as a solid diamond and can be used to denote possible alternatives in satisfying a relationship \cite{b33}. 
\newline
Figure \ref{fig2} shows a sample assurance case pattern adapted from \cite{b61}. That pattern is represented in GSN.

\section{METHODOLOGY}
\label{sec:4}
\begin{figure}[hbt!]
    \centering
    \includegraphics[width=1\linewidth]{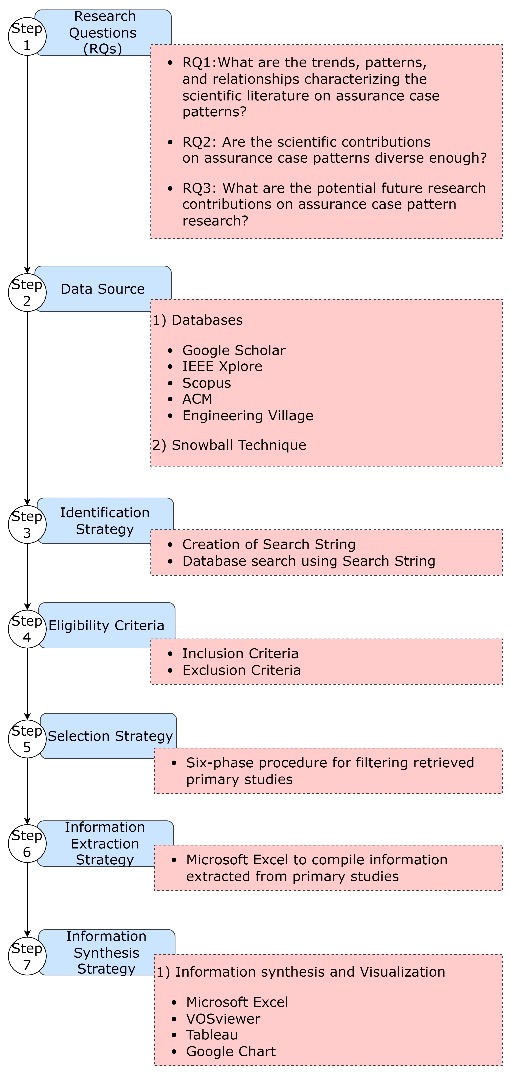}
    \caption{Methodology framework for our bibliometric analysis}
    \label{fig3}
\end{figure}

To conduct our bibliometric analysis, we relied on SEGRESS (Software Engineering Guidelines for Reporting Secondary Studies) \cite{b44}. SEGRESS is an adaptation of PRISMA (Preferred Reporting Items for Systematic Reviews and Meta-Analyses) 2020 \cite{b43} for the software engineering field. We also relied on common bibliometric analysis techniques \cite{b14}. Figure \ref{fig3} presents a high-level overview of our methodology. We further describe the steps of our methodology in the remainder of this section. 

\subsection{Research questions}
Our bibliometric analysis aimed at investigating the three research questions (RQs) below:\newline
\textbf{RQ1: What are the trends, patterns, and relationships characterizing the scientific literature on assurance case patterns?} 

This question aims to explore the research evolution, development trend, and significance of assurance case patterns over time. In this regard, we identify and visualize the most cited keywords and publications to unravel the factors for the adoption of assurance case patterns within the last two decades. In this research question, we also identify the most impactful conferences, journals, and workshops with the highest contributions to assurance case patterns. \newline
\textbf{RQ2: Are the scientific contributions on assurance case patterns diverse enough?} 

This question aims to map out the diversity of the scientific  landscape for the publications on assurance case patterns. We investigate the diversity in the collaboration network, author affiliation, and countries of the most active and prolific global researchers in the field of assurance case patterns. \newline
\textbf{RQ3: What are the potential future research contributions on assurance case pattern research?} 

This question aims to provide insights into emerging trends, possible future research directions, and areas of priority for the application of assurance case patterns.

\subsection{Information sources}

\subsubsection{Database-driven Search}

Our search for primary studies relating to assurance case patterns was conducted on five well-known scientific search engines; IEEE Xplore \cite{b67}, Scopus \cite{b69}, ACM Digital Library \cite{b68}, Engineering Village \cite{b70}, and Google Scholar. To automatically search Google Scholar, we used the publish or perish tool \cite{b45}. 

\subsubsection{Snowballing Technique}

To ensure high coverage of primary studies related to assurance case patterns during our search process, we also performed snowballing \cite{b60}.
We utilized the Connected Papers tool \cite{b46} to automatically perform the forward and backward snowballing technique. We used the primary studies found during our database-driven search as the start set for the snowballing process.

\subsection{Identification strategies}

To ensure the retrieval of all relevant studies, we first conducted a manual search for papers related to assurance case patterns to become acquainted with the keywords commonly used by researchers in this field. Using the insights gained from this initial search, we formulated the search strings provided in the text box below, and applied them across five scientific databases.

\smallskip
\noindent\fbox{%
    \parbox{\linewidth}{%

\textbf{1: }
\textit{``safety case pattern'' \textcolor{red}{OR} ``security case pattern'' \textcolor{red}{OR} ``assurance case pattern'' \textcolor{red}{OR} ``argument pattern''}

\smallskip

\textbf{2: }
\textit {(``safety case'' \textcolor{red}{OR} ``security case'' \textcolor{red}{OR} ``assurance case'') \textcolor{red}{AND} (``pattern'')}
}
}
\smallskip
\newline
The first portion of the search string aimed at increasing the breadth and extending the area of the search to include all occurrences of safety case pattern, security case pattern, assurance case pattern, and argument pattern. The second portion of the search aimed in turn at limiting the search string to patterns for common functional requirements. These search strings were used in the advanced search of the five different databases.

\subsection{Inclusion and exclusion criteria}

Table \ref{tab1} reports the inclusion and exclusion criteria we utilized to select primary studies. We carefully formed these criteria when designing our review protocol. This made it easy to filter the retrieved primary studies. For the inclusion criteria, we considered studies focusing on the representation, instantiation, creation, assessment, reuse, and formalization of assurance case patterns. We also considered studies that are peer-reviewed and studies written in English as this is the common language among authors of this paper. To limit outdated studies and maintain relevancy, we only considered studies published within the last two decades (January 2003 to October 2023). Furthermore, we did not consider short papers (less than four pages) as well as not peer-reviewed papers. We made that choice to ensure that studies utilized for the bibliometric analysis had enough material relevant for our analysis and were of great quality.

\begin{table*}[t!]
  \caption{Inclusion and exclusion criteria}
  \label{tab1}
  \begin{tabular}{p{8cm} p{8cm}} 
    \toprule
    Inclusion criteria&Exclusion criteria\\
    \midrule
    1. Conference papers, Workshop papers, Journal papers, Peer reviewed papers & 1. Books, posters, book chapters, grey literature (e.g., Computing Research Repository (CoRR) in Arxiv, SSRN), theses, tutorials, technical reports, papers with less than 4 pages \\
    2. The study is written in English, the study has been published in the last 20 years (between 2003 and October 2023)& 2. Studies not in English language \\
    3. Papers discussing how Assurance case patterns are represented. & 3. Papers that do not propose techniques to design, model, represent, or generate assurance cases using patterns \\
    4. Papers describing how patterns are used for assurance case creation.& 4. Incomplete papers (only abstract is available)\\
    5. Papers applying a case study in the use of assurance case generated from an assurance case pattern.& 5. Non-context assurance, secondary studies (e.g., reviews, systematic literature reviews, systematic mapping studies, surveys)\\
    6. Papers focusing on assessment, reuse, formalization, automation, maintenance, and instantiation of assurance case patterns. & 6. Papers with paywall preventing free access to the paper and emailing the authors of the paper to get it was unfruitful \\
    7. Papers focusing on patterns used in assurance cases & 7. Papers with Assurance case patterns, but no description of how the assurance case patterns were utilized\\
  \bottomrule
\end{tabular}
\end{table*}

\subsection{Selection strategies}

\subsubsection{Database-driven}

Table \ref{tab2} shows the strategy we utilized to narrow down the primary studies for our bibliometric analysis.
After we completed our database-driven search, we proceeded with a six-phase approach to filter out unrelated studies from our retrieved primary studies. In the first phase, we imported all retrieved studies from the five different databases into a reference manager tool called EndNote \cite{b47}. In the second phase, we filtered out studies that were not peer-reviewed and studies without titles. The third phase involved using the "find duplicates" feature in Endnote to automatically remove duplicate studies from the different database results. We filtered out studies based on titles, keywords, abstracts, venues, inclusion, and exclusion criteria in the fourth phase. Furthermore, in the fifth phase, we filtered out studies by reading the introduction, and conclusion of the studies and using the inclusion and exclusion criteria. Finally, in the last phase, we eliminated unrelated studies based on full-text reading, inclusion, and exclusion criteria.

\subsubsection{Snowballing}

When completing the snowballing process, we also followed the same six-phase approach listed in Table \ref{tab2} for filtering out primary studies with a slight modification in the duplicate removal step. 
In the first snowballing iteration, we used the connected paper tool \cite{b46} to generate a graph of related papers for each primary study found in the database-driven search. This is then followed by the six-phase approach listed in table \ref{tab2}. 

During phase three of the first snowballing iteration, we removed all duplicate references in this phase and also compared them with references from phase one of the database-driven search to identify and further remove any duplicate studies.
In the subsequent snowballing iteration, we used the result of the last filtering phase (phase six) in the previous snowballing iteration as a start set and followed the six-phase approach listed in table \ref{tab2} for selecting primary studies. 

During phase three of the subsequent snowballing iteration, we removed all duplicate references in this phase and also compared them with references from phase one of the database-driven search and the previous snowballing iterations phase one to identify and further remove any duplicate studies.

The listed phases in table \ref{tab2} for primary study selection were executed by a sole researcher. To minimize potential bias, a second researcher independently and randomly sampled studies in each phase for validation. Furthermore, regular meetings were held to address and resolve any disagreements between the two researchers.

\subsection{Data extraction strategies}

In our bibliometric analysis, we used Excel sheets to compile data from the retrieved primary studies, including details such as publication year, authors, titles, venue, and the number of citations. The citation count for each primary study was manually obtained from Google Scholar. Additionally, We extracted the primary studies in the EndNote library in RIS (Research Information Systems) format and used them as input to VosViewer \cite{b15} for data synthesis. 

\begin{table}[t!]
  \caption{Primary Studies Selection Strategy}
  \label{tab2}
  \begin{tabular}{p{1cm} p{7cm}} 
    \toprule
    Phase&Action Performed\\
    \midrule
    1. &Importing in EndNote the references of studies found in the searched databases\\
    2. &Cleaning references (e.g., with no title, of study type not covered, not peer-reviewed)\\
    3. &Removing duplicates (i.e., identical references coming from different databases)\\
    4. &Excluding studies based on the titles, keywords, abstracts, venues, and inclusion and exclusion criteria\\
    5. &Excluding studies based on the introductions and conclusion, and inclusion and exclusion criteria\\
    6. &Excluding studies based on full-text reading, and inclusion and exclusion criteria\\
  \bottomrule
\end{tabular}
\end{table}

\subsection{Synthesis strategies}

In our bibliometric analysis, we employed the widely used VosViewer tool \cite{b15} to synthesize data extracted from primary studies. We used VosViewer to facilitate the analysis, synthesis, and visualization of trends and patterns in published research works related to assurance case patterns. Furthermore, for the generation of additional charts, we utilized Microsoft Excel, Google Charts \cite{b48}, and Tableau \cite{b16}

\section{RESULTS}
\label{sec:5}

\begin{figure*}[htbp]
    \centering
    \includegraphics[width=1\linewidth]{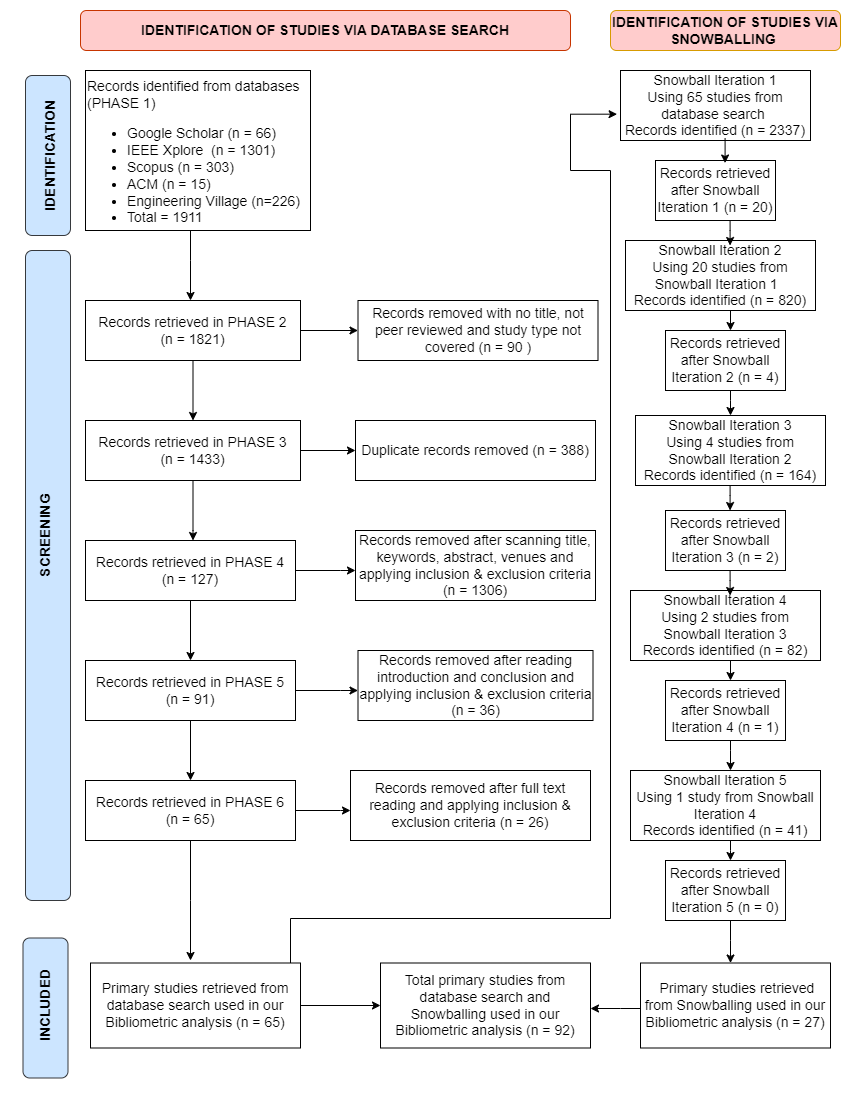}
    \caption{PRISMA flow diagram illustrating the identification and selection of primary studies}
    \label{fig4}
\end{figure*}

Figure \ref{fig4} shows our PRISMA flow chart diagram. The total number of primary studies identified, filtered out, and included using our database-driven search and snowballing technique is depicted in this diagram. We identified and utilized a total of \textbf{92} primary studies to answer our proposed research questions in the subsequent subsections.

Table \ref{tab4} in the Appendix reports the details about our list of primary studies. That information consists of:  the authors of the study, its publication year, its title, its venue acronym, and the search method (i.e. snowballing, database-driven search) we used to select that study. 

\subsection{RQ1:{What are the trends, patterns, and relationships characterizing the scientific literature on assurance case patterns?}}

\subsubsection{\textbf{Annual Distribution of Publications in Assurance case pattern}}
\begin{figure}[htbp]
    \centering
    \includegraphics[width=1\linewidth]{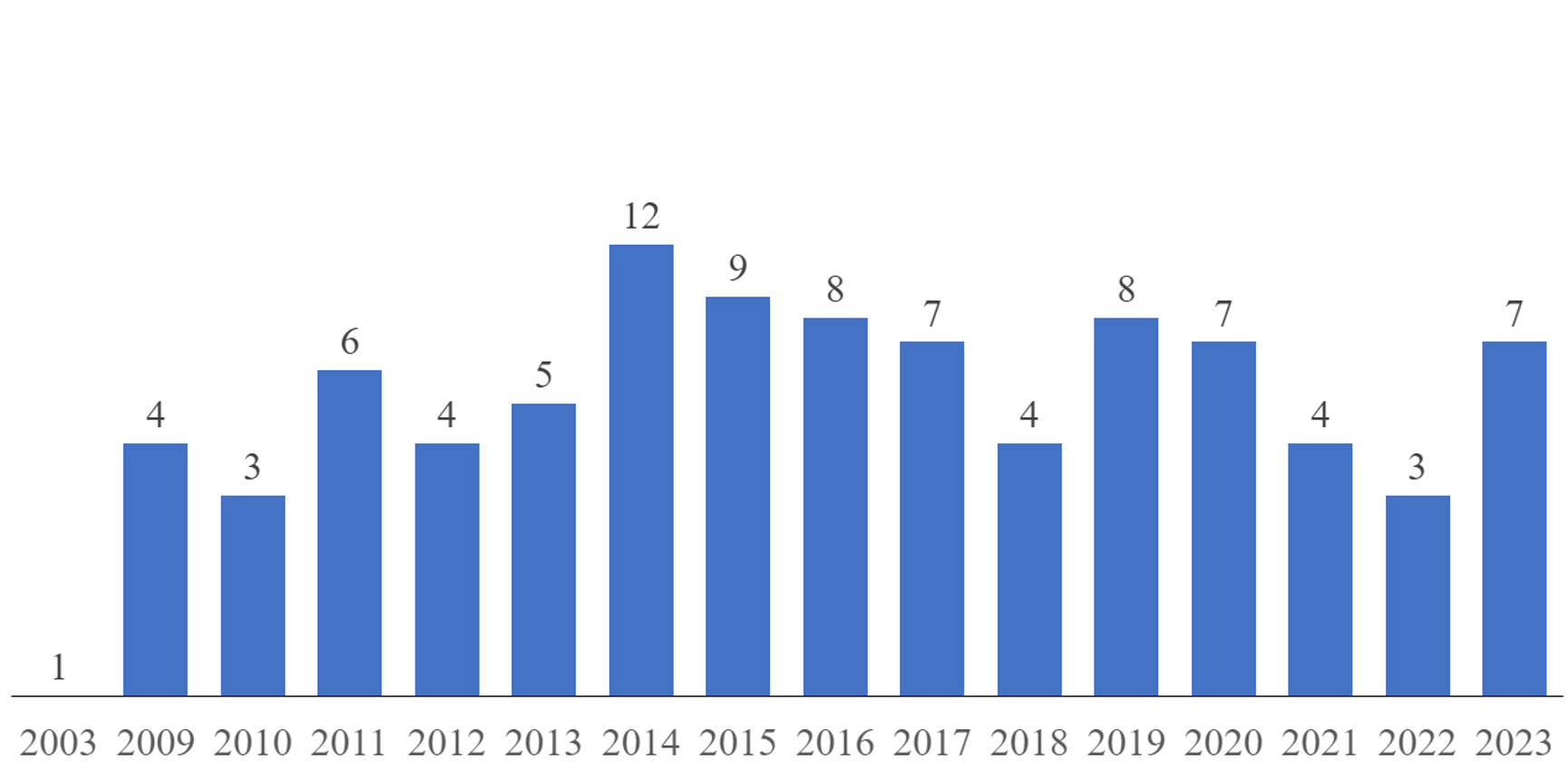}
    \caption{Publication-year distribution}
    \label{fig5}
\end{figure}
Figure \ref{fig5} reports the number of publications and their distribution over time within the last two decades. From this distribution, the number of publications per year before 2009 is very low. An explanation for this might be due to the novelty of assurance case patterns and the existence of less complex interconnected systems. We observed that the number of publications started to increase in 2009 which could be attributed to the introduction and popularization of model-based systems engineering (relevant in the development and conceptualization of assurance case patterns) in 2007 by the International Council on Systems Engineering (INCOSE). 

Specifically, the number of publications fluctuated between 2009 and 2013 leading to an average of four publications between 2009 (inclusive) and 2013 (inclusive). However, between 2014 and 2018, there was an average of eight publications per year. 
A notable observation in the distribution is the peak of 12 publications in 2014 which could be attributed to the increase in the adoption of templates that encapsulate and support the re-use of evidence and arguments across multiple domains to reduce cost and time in assuring the safety of a system and their compliance with safety standards. 

In 2021 and 2022, there was a decrease in the number of publications which may be due to the COVID-19 pandemic that disrupted activities across the world and a decline in opportunities for collaboration. It is crucial to point out that our search for primary studies ended in early October 2023 which means that our search might miss a few studies published between early October 2023 and December of 2023. However, our publication-year distribution for 2023 shows an increase in the number of publications compared to the two previous years which signifies ongoing interest and relevance of research on assurance case patterns.

\subsubsection{\textbf{Distribution of Publication by Venue}}
\begin{figure}[htbp]
    \centering
    \includegraphics[width=1\linewidth]{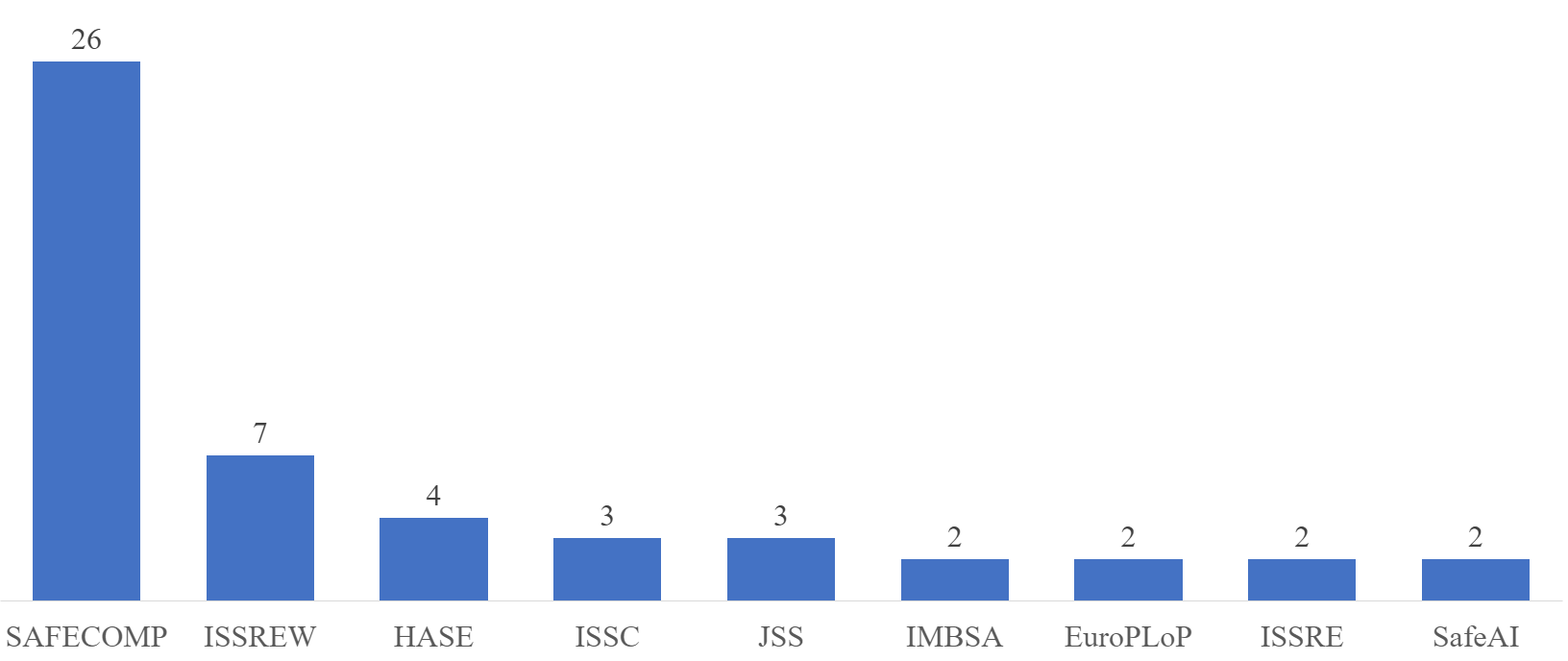}
    \caption{Publication venue distribution (January 2003 - October 2023)}
    \label{fig6}
\end{figure}

Gaining knowledge about the focus of each conference, journal, or workshop is not only important for assisting new researchers decide the right venue for sharing their research, but also for identifying which venue to search and explore when performing a literature review. To this end, Figure \ref{fig6} provides information about the most influential hotspot venues for publishing articles on assurance case patterns. That chart shows that SAFECOMP (International Conference on Computer Safety, Reliability, and Security) is the most preferred venue since it yields the highest number of publications. This corroborates with the dedication of SAFECOMP in the promotion of research focusing on the dependability and reliability of safety-related and safety-critical systems.

A noteworthy observation is ISSREW (International Symposium on Software Reliability Engineering Workshops) which is a workshop of ISSRE (International Symposium on Software Reliability Engineering) ranks second with seven publications which shows that new researchers with viable ideas can publish their research in workshops and gather feedback. Next is the HASE (International Symposium on High-Assurance Systems Engineering) with four publications. However, the last occurrence of this conference was in 2019. Also, SafeAI (The AAAI’s Workshop on Artificial Intelligence Safety) being part of the top nine venues shows an ongoing focus on ethics assurance of machine learning systems and safety assurance of systems with machine learning components.

\subsubsection{\textbf{Top publications by citations}}
To identify some of the top publications that have a significant impact in the field of assurance case pattern, we utilized Google Scholar \cite{b71} to retrieve the count of citations for each of the primary studies used in our bibliometric analysis.
Table \ref{tab3} shows the top 16 most influential publications based on the number of citations retrieved from Google Scholar \cite{b71} as of December 12, 2023.  All citations per publication in the table exceed 30, with an average citation value of 64. 

The publications by Hawkins et al. \cite{b49} and Hawkins et al. \cite{b50} led the list with 206 and 116 total citations respectively and published by the same first author. This is followed by Denney and Pai \cite{b51} and Denney and Pai. \cite{b39} with 115 and 69 total citations and published by the same first author. 
Hawkins et al. \cite{b49} proposed a separation of concerns approach for creating a clear assured safety argument that distinguishes the safety argument from its associated confidence arguments to ensure easy identification and mitigation of assurance deficits. In their work, they proposed ACPs to demonstrate sufficient confidence that assurance deficits have been identified. mitigated and that any residual deficits are acceptable. 
Also, Hawkins et al. \cite{b50} proposed a model-based assurance approach that uses a weaving model to link argument patterns with information extracted from design models, analysis models, and development models of a system. This approach aimed to automate the generation and analysis of assurance cases, improve traceability between the assurance case and the design and analysis models, and support the coevolution of the system design and the assurance case. 

Denney and Pai \cite{b51} developed AdvoCATE a robust multi-functional tool-set for the automatic creation of assurance cases, instantiating argument patterns, support for integrating evidence from formal methods into assurance cases, and support for hierarchy and modular organization of argument structures.  
The publications by Hawkins \cite{b49, b50} and Denney \cite{b51, b39} perform well in terms of the number of citations which shows its strong relevance and high influence in assurance case patterns research. Thus, new researchers should pay more attention to these publications. 

A glance of Table \ref{tab3} shows the terms \textbf{\textit{``Patterns''}}, \textbf{\textit{``model-based approach''}}, \textbf{\textit{``automotive''}}, \textbf{\textit{``machine learning''}}, and \textbf{\textit{``formal''}} as the most common keywords in the titles of these publications. This suggests the use of assurance case patterns in the automotive industry, the need for ACPs to provide argument structure for systems with machine learning components, the formalization of ACPs, and the need for Model-based assurance cases using SACM.

\begin{table*}
    \centering
    \caption{Top 15 publications based on the number of citations on Google Scholar as of December 12, 2023}
    \begin{tabular}{p{1cm} p{3cm} p{11cm} p{1.5cm}}
    \toprule
     No & Author & Title & \# Citations\\
     \midrule
         1&Hawkins et al. (\citeyear{b49})&A New Approach to Creating Clear Safety Arguments&206 \\
         2&Hawkins et al. (\citeyear{b50})&Weaving an Assurance Case from Design: A Model-Based Approach  &116 \\
         3&Denney \& Pai.  (\citeyear{b51})&Tool support for assurance case development&115 \\
         4&Denney \& Pai.  (\citeyear{b39})&A formal basis for safety case patterns&69 \\
         5&Yamamoto et al. (\citeyear{b39})&An evaluation of argument patterns to reduce pitfalls of applying assurance case&63 \\
         6&Wei et al. (\citeyear{b35})&Model Based System Assurance Using the Structured Assurance Case Metamodel&62 \\
         7&Weaver et al. (\citeyear{b63})&A Pragmatic Approach to Reasoning about the Assurance of Safety Arguments&50 \\
         8&Ayoub et al. (\citeyear{b64})&A Systematic Approach to Justifying Sufficient Confidence in Software Safety Arguments&47 \\
         9&Ayoub et al. (\citeyear{b57})&A safety case pattern for model-based development approach&46 \\
         10&Wagner et al. (\citeyear{b6})&A Case Study on Safety Cases in the Automotive Domain: Modules, Patterns, and Models&45 \\
         11&Picardi et al. (\citeyear{b3})&A Pattern for Arguing the Assurance of Machine Learning in Medical Diagnosis Systems&41 \\ 
         12&Palin et al. (\citeyear{b5})&Assurance of Automotive Safety - A Safety Case Approach&40 \\
         13&Burton et al. (\citeyear{b7})&Confidence Arguments for Evidence of Performance in Machine Learning for Highly Automated Driving Functions&36 \\
         14&Hawkins et al. (\citeyear{b41})&A Systematic Approach for Developing Software Safety Arguments&34 \\
         15&Sljivo et al. (\citeyear{b65})&Generation of Safety Case Argument-Fragments from Safety Contracts&32 \\
         16&Matsuno et al. (\citeyear{b66})&Parameterised Argument Structure for GSN Patterns&32 \\
      \bottomrule
    \end{tabular}

    \label{tab3}
\end{table*}

\subsubsection{\textbf{Most cited/influential keyword analysis}}

\begin{figure*}[htbp]
    \centering
    \includegraphics[width=1\linewidth]{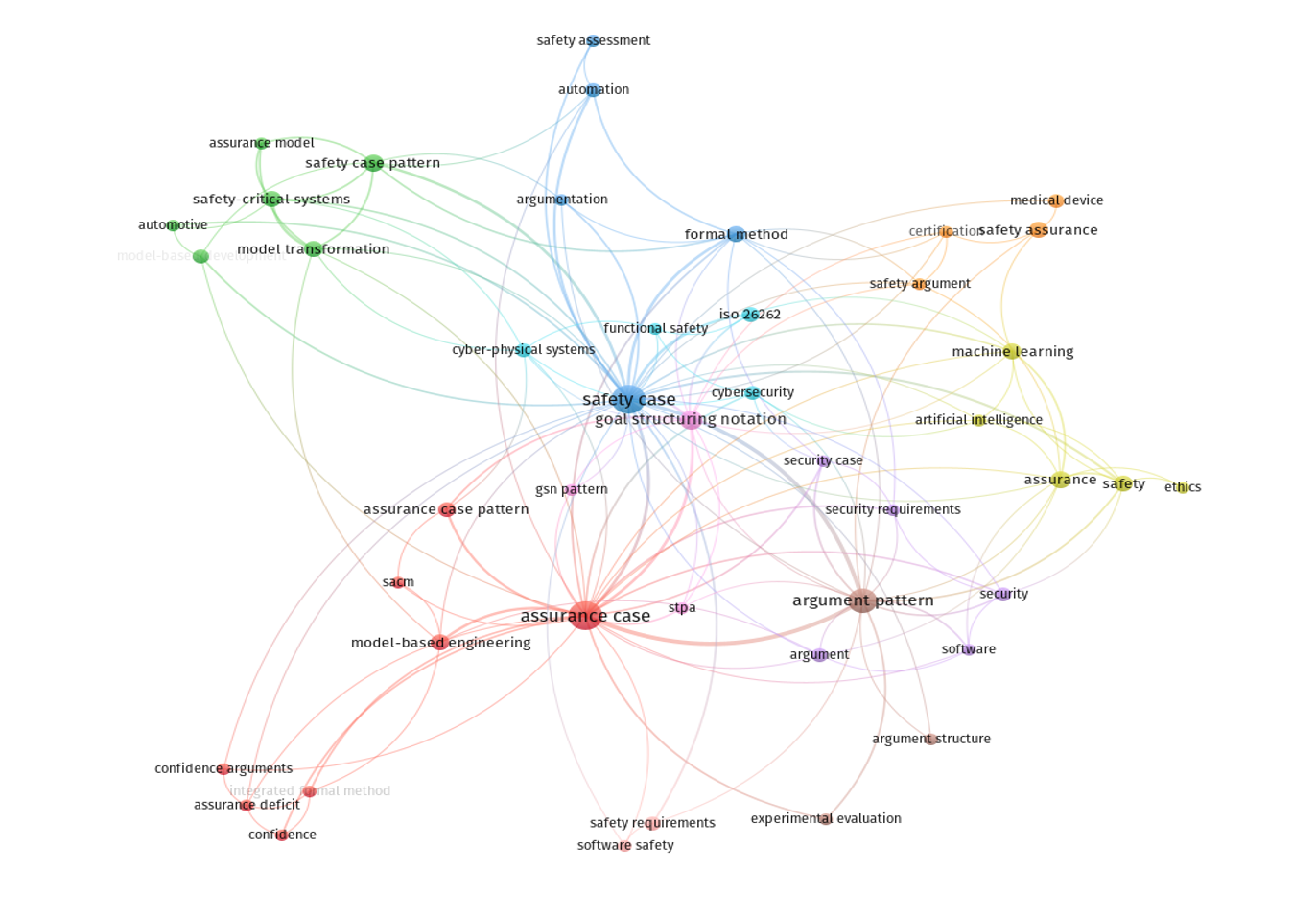}
    \caption{Author keyword analysis}
    \label{fig7}
\end{figure*}

To understand the knowledge structure and conceptual structure of main topics of research interest across the field of assurance case patterns, we used VosViewer to generate the keyword network that Figure \ref{fig7} illustrates. VosViewer  synthesizes frequently used keywords and analyze the keyword co-occurrence network across the field of ACP. Based on our dataset of 92 papers in ACP and a total of 224 keywords retrieved from these papers, we set the minimum number of occurrences of a keyword as two in VosViewer and eliminated some inconsistencies by merging semantically identical keywords like \textbf{\textit{"GSN"}} and \textbf{\textit{"Goal Structuring Notation"}}. Based on these criteria, only 45  keywords met this threshold. We used these 45 keywords to generate the keyword network. 

In a keyword network, a node represents a keyword, and the bigger the node, the more frequently the keyword appears. A link between two nodes means the co-occurrence of two keywords. Similarly, each color in our keyword network represents a cluster of keywords with a similar theme, and there are 10 clusters in Figure \ref{fig7}. 

In Figure \ref{fig7}, we can see the terms \textbf{\textit{``safety case''}}, \textbf{\textit{``assurance case''}}, \textbf{\textit{``goal structuring notation''}}, \textbf{\textit{``argument pattern''}}, \textbf{\textit{``safety case pattern''}}, and \textbf{\textit{``assurance case pattern''}} as the most prominent keywords taking notable potions in the chart. This is not surprising as the majority of these keywords are present in the search strings used to retrieve relevant primary studies for this study. In addition, these keywords form the fundamental background terms in the field of ACP.
The presence of Goal structuring notation (GSN) --a graphical argumentation notation \cite{b33} -- among these prominent keywords, shows the widespread use of GSN in the representation of assurance cases and assurance case patterns. 

In the keyword network, the \textbf{\textit{``SACM''}} keyword stands for Structured Assurance Case Metamodel \cite{b49}. SACM is a specification defined by the Object Management Group (OMG) for representing structured assurance cases \cite{b49}. Wei et al \cite{b35}, provided an exposition on the usage and robust features of SACM, and the relationship that exists between SACM and other notations like CAE (Claim-Argument-Evidence) \cite{b8} and GSN \cite{b25} for representing assurance cases. Also, Selviandro \cite{b53} proposed an extension of SACM in SACM notation (SACMN) which is a syntax of visual vocabularies and compositional rules for representing assurance cases and assurance case patterns. Both Wei et al. \cite{b35} and Selviandro \cite{b54} aimed to simplify the complexity of SACM and drive the adoption of its use for representing assurance cases and assurance case patterns. 

Also, other prominent sets of keywords include \textbf{\textit{``model-based engineering''}}, \textbf{\textit{``model transformation''}}, \textbf{\textit{``model-based development''}}, \textbf{\textit{``assurance model''}}, and \textbf{\textit{``automation''}}. These keywords provide evidence of the direct link that exists between assurance case patterns and system design models. Numerous studies in assurance case patterns \cite{b35,b50,b52}  have proposed different approaches to automate the generation of assurance cases from assurance case patterns by instantiating ACPs using extracted information from system design models. 
Wagner et al. \cite{b6} investigated the structure and application of models (functional models, platform models, environment model) as “information and requirement sources to guide the construction of safety cases using ACP along the architecture of a system" in the automotive domain. 
In another recent study, Hartsell et al. \cite{b52} also proposed an automated method for constructing assurance cases for complex CPS based on the instantiation of ACP using information extracted directly from a set of interconnected models. Their focus was on generating assurance cases that maintain and ensure traceability for each argument and evidence used in the AC back to the artifact it was sourced from in the system design model. 

Keywords such as \textbf{\textit{``Safety critical systems''}}, \textbf{\textit{``Automotive'', ``Medical device''}}, \textbf{\textit{``cyber-physical systems''}} embody some application domains and type of systems that the literature on assurance case patterns focuses on. 
In the automotive domain, assurance case patterns in the form of safety case patterns have been proposed to generate safety cases for cruise control systems \cite{b6}, braking systems \cite{b54,b55}, airbag systems~\cite{b8}. 
In the medical domain, we have patterns that provide a reusable argument structure for safety compliance of Generic Patient Controlled Analgesic (GPCA) \cite{b56,b57}.
Also, keywords like \textbf{\textit{"ISO 26262"}}, \textbf{\textit{"certification"}} show that ACP can be used to support the compliance of different industry standards and certification of systems.

ACP targets different functional requirements to support their assurance. The keywords \textbf{\textit{``functional safety''}}, \textbf{\textit{``security requirement''}}, \textbf{\textit{``safety requirement''}}, \textbf{\textit{``Software safety''}}, \textbf{\textit{``security''}}, \textbf{\textit{``cybersecurity''}} provide credence to some of the requirements that research in ACP targets. From these keywords, it is evident that safety and security currently rank highest as the requirements supported by ACP. 
In addition, keywords like \textbf{\textit{``safety assessment''}}, \textbf{\textit{``confidence arguments''}}, \textbf{\textit{``assurance deficit''}}, \textbf{\textit{``confidence''}}, \textbf{\textit{``integrated formal method''}}, \textbf{\textit{``formal methods''}} in the red cluster of Figure \ref{fig7}  represent another area in the field of ACP that is concerned with formalization of ACP and the use of formal methods in system assurance to eliminate assurance deficits. 

Finally, keywords in the yellow cluster are, \textbf{\textit{``Machine learning''}}, \textbf{\textit{``artificial intelligence''}}, \textbf{\textit{``assurance''}}, \textbf{\textit{``safety''}}, \textbf{\textit{``ethics''}}. These keywords reflect the emergence of new assurance case patterns to support various emerging machine learning-enabled systems and evolving system requirements, including those related to ethics. 

\smallskip
\noindent\fbox{%
    \parbox{\linewidth}{%
    \smallskip
    In summary, for \textbf{RQ1}, The number of publications in ACPs have fluctuated in the last two decades with a peak of 12 publications in 2014. However, in 2023, our publication-year distribution shows an increase in the number of publications compared to the two previous years which signifies ongoing interest and relevance of research in assurance case patterns. Also, SAFECOMP emerges as the top choice venue for the publication of studies in ACPs. Two studies from both Richard Hawkins \cite{b49, b50} and Ewen Denney \cite{b51, b39} emerge as the top cited studies from our list of primary studies.
Finally, keywords like \textbf{“Machine learning”}, \textbf{“artificial intelligence”}, \textbf{“ethics”} have emerged which indicate the emergence of new assurance case patterns to support various emerging machine learning-enabled systems and evolving system requirements, including those related to ethics.

}}

 \subsection{RQ2: {Are the scientific contributions on assurance case patterns diverse enough?}}

\subsubsection{\textbf{Distribution of author's affiliation}}
\begin{figure*}[htbp]
    \centering
    \includegraphics[width=1\linewidth]{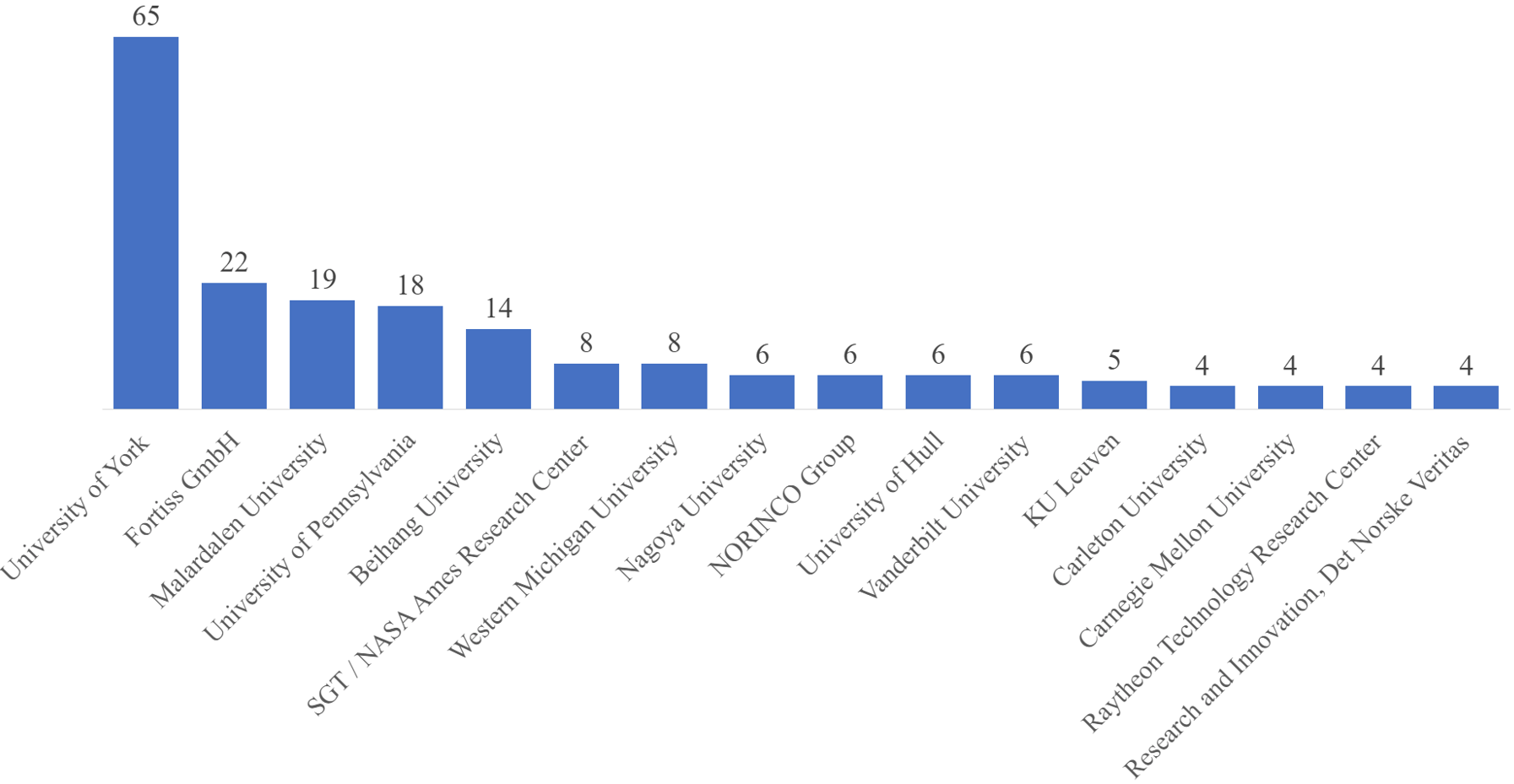}
    \caption{Author's affiliation distribution}
    \label{fig8}
\end{figure*}

Figure \ref{fig8} shows the top institutions with the highest number of authors contributing to the field of assurance case pattern. The University of York ranks first with 65 authors followed by Fortiss Gmbh research institute in Germany with 22 authors. Ranked third in this distribution is Malardalen University in Sweden with 19 authors. This is closely followed by the University of Pennsylvania with 18 authors and Beihang University in China in fifth position with 14 authors. In addition, a noteworthy mention is the SGT/NASA Ames Research Center with 8 authors ranked in the sixth position. 

This distribution corresponds with Table \ref{tab3} as six publications from the top 15 publications have their first authors affiliated with the University of York which shows the commitment and dedication of the University of York to the field of assurance case patterns. 
Also, A key observation is that the USA has the highest number of diverse institutions within this chart. However only one of its institutions emerged in the top five for author’s affiliation. 

\subsubsection{\textbf{Distribution of author's country of affiliation}}

\begin{figure*}[htbp]
    \centering
    \includegraphics[width=1\linewidth]{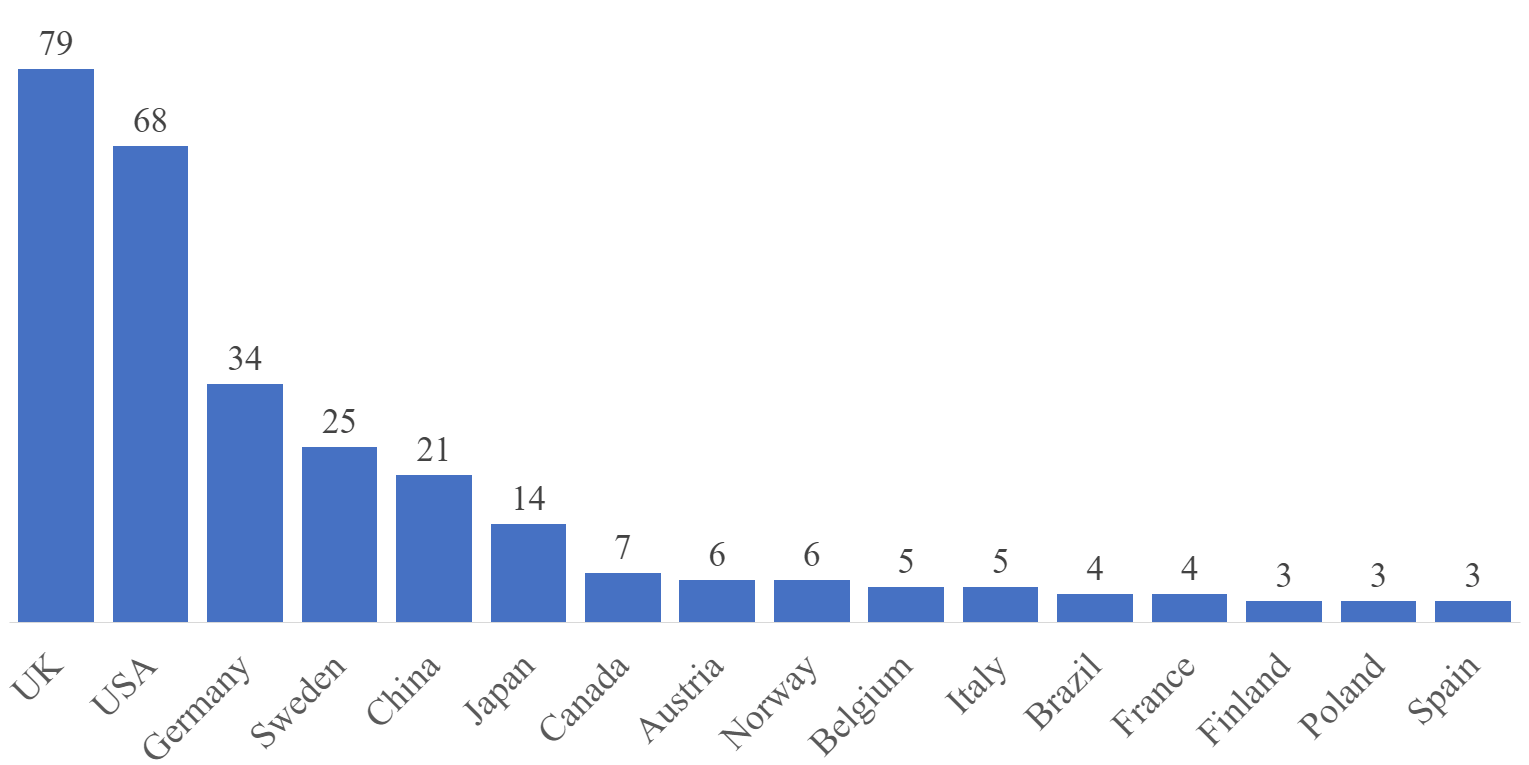}
    \caption{Distribution of author's country}
    \label{fig9}
\end{figure*}

Different scholars worldwide have tried to expand the body of knowledge in ACP. To better understand this research diversity across countries of the world, we generate a chart to show the distribution of country of affiliations of authors in our study. 
In figure \ref{fig9}, The United Kingdom emerged first with the highest number of authors in the assurance case pattern. This can be attributed to the presence of the University of York a pacesetter, pioneer, and leader in the field of assurance case patterns in the United Kingdom. 

The next prominent country in the chart is the United States of America (USA) with 68 authors. This aligns with figure \ref{fig8}, which shows the USA has the highest number of institutions within the chart. Furthermore, agencies like the Food and Drug Agency (FDA) within the USA that are concerned with the safety of medical devices, and NASA is concerned with the safety of space shuttles and unmanned vehicles might have contributed to the USA being highly ranked in this chart. 

Also, the presence of Germany as the third-ranked country could be due to Fortis Gmbh (a research center that supports safety and security in software and system development). In addition, popular car brands like BMW, and Volkswagen are based in Germany, and from their perspectives, ensuring the safety of automotive vehicles is important. 

\subsubsection{\textbf{Geographical Landscape of Assurance case Pattern}}

\begin{figure*}[htbp]
    \centering
    \includegraphics[width=1\linewidth]{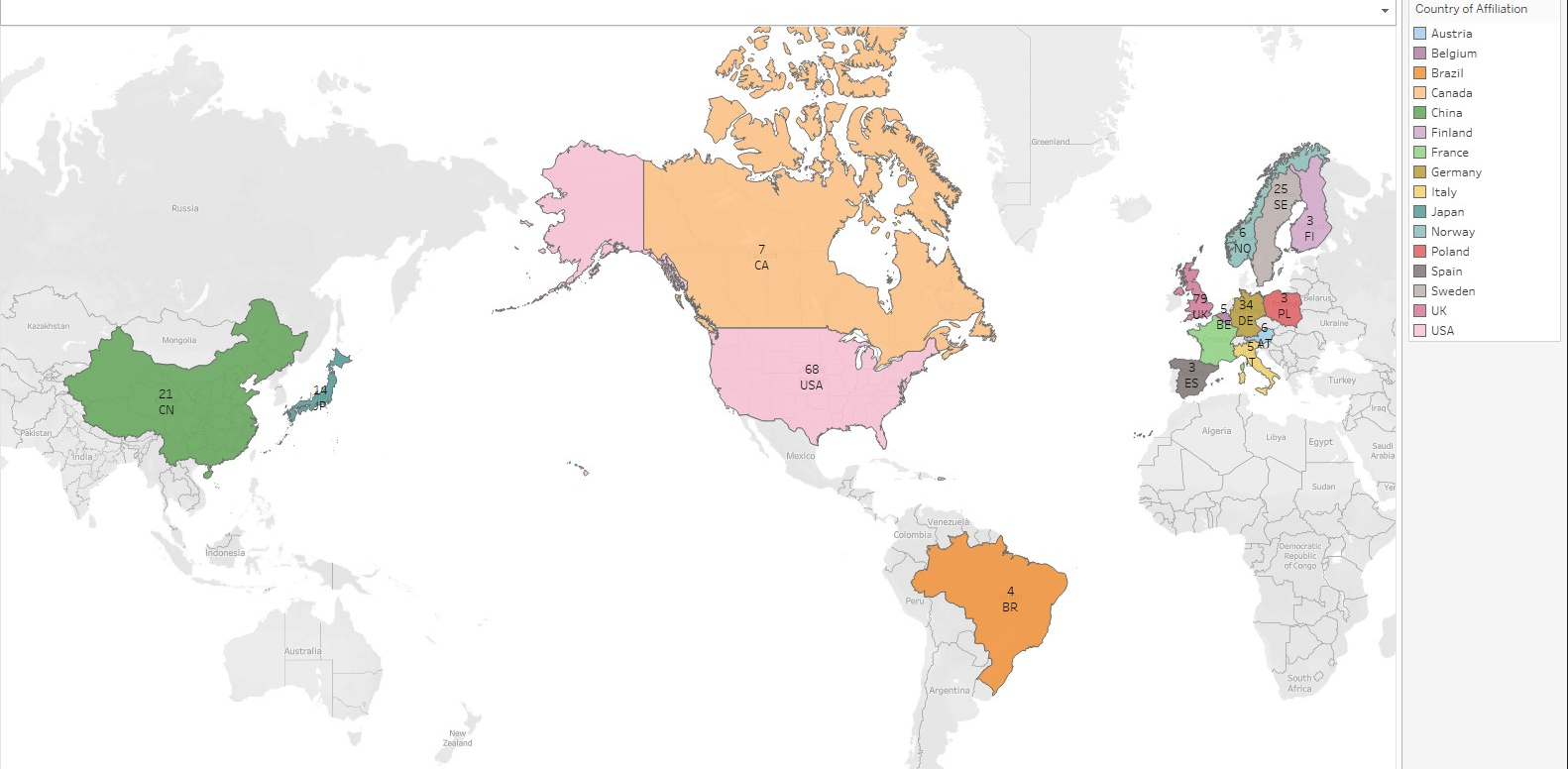}
    \caption{World map showing country of affiliation}
    \label{fig10}
\end{figure*}

This section focuses on the geographical distributions of assurance case patterns based on the authors' country of affiliation. This analysis allows us to better understand the geographic distribution of researchers contributing to the field of assurance case patterns. 
We used Tableau \cite{b16} to create the world map distribution \footnote{In this Figure, UK = United Kingdom; USA = United States of America; DE = Germany; SE = Sweden; CN = China; JP = Japan; CA = Canada; AT = Austria; NO = Norway; BE = Belgium; IT = Italy; BR = Brazil; FR = France; FI = Finland; PL = Poland; ES = Spain} in Figure \ref{fig10}. From the world map, each highlighted area represents an author’s country of affiliation. 

European countries are the main contributors in the field of assurance case patterns, having 11 countries represented in the top 16 countries. The three other representative continents are North America, Asia, and South America. Only two countries (i.e. Canada and the United States of America)  are represented in North America while China and Japan are representatives from Asia. There is only one representation for South America which is Brazil. 

The reason for the absence of some countries in Asia, South America, Europe (e.g., Russia) and North America, and the total absence of countries from Africa and Oceania could be due to language barrier. This makes sense since the studies we analyzed in our bibliometric analysis are exclusively written in English. This absence could also be due to the possibility that researchers with nationalities of these countries are affiliated with organizations in foreign countries prominent in the field of assurance case patterns. 

Furthermore, the strong prominence of European countries in the field of assurance case patterns can be attributed to the common standards, policies, and rules that apply automatically and uniformly to all European Union countries. Based on these findings, the geographical location of countries tends to influence the degree of research and development of the field of assurance case patterns. Hence, there is a need for active collaboration and cooperation among researchers from all countries or continents to speed up the  development of the field of assurance case patterns especially in under-represented regions.

\subsubsection{\textbf{Analysis of co-Author network}}

\begin{figure*}[htbp]
    \centering
    \includegraphics[width=1\linewidth]{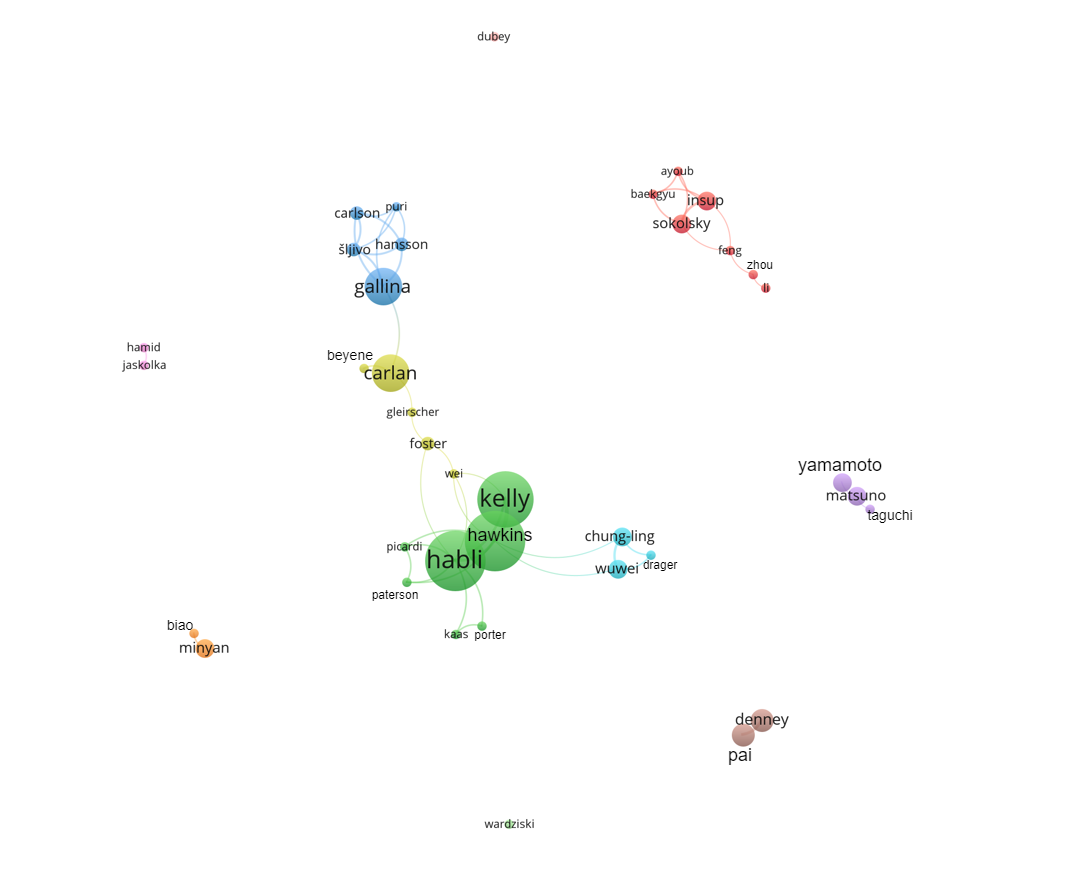}
    \caption{Co-author network}
    \label{fig11}
\end{figure*}

\begin{figure*}[htbp]
    \centering
    \includegraphics[width=1\linewidth]{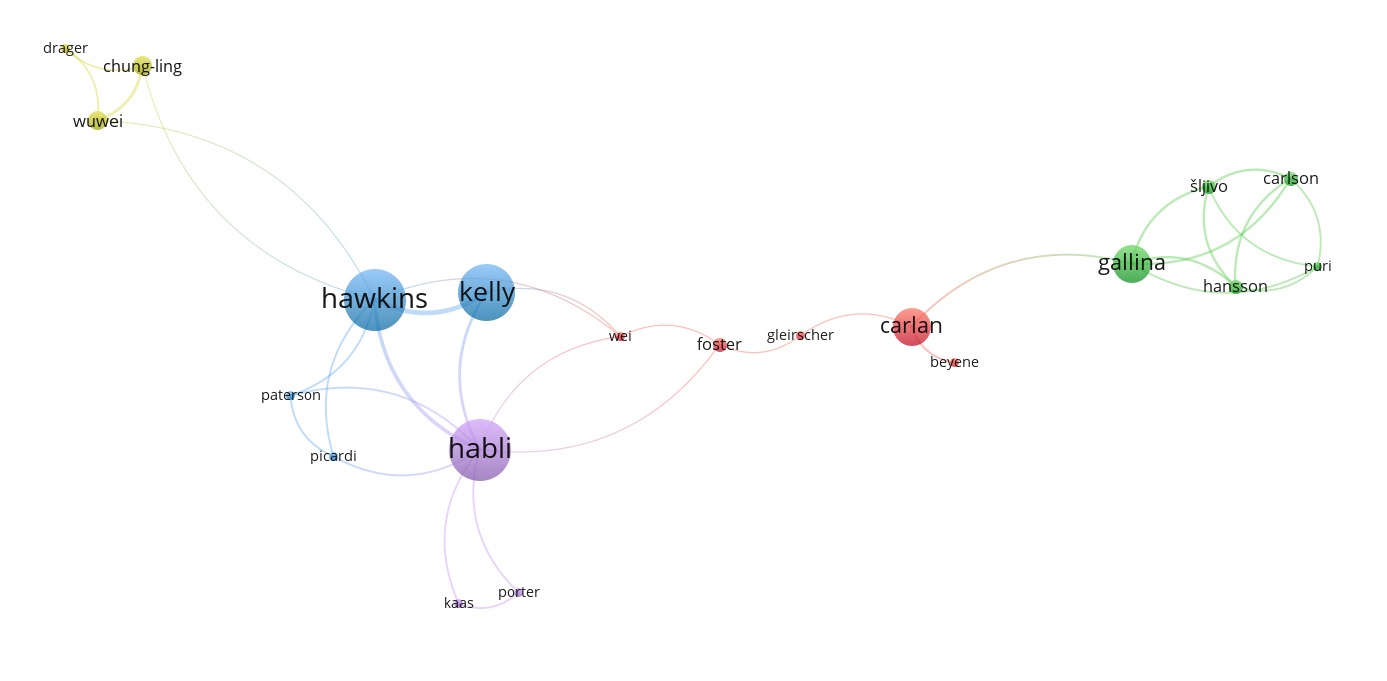}
    \caption{Connected Co-author network}
    \label{fig12}
\end{figure*}

To analyze the authors in our dataset of primary studies by publications made together, we created two co-author networks using the visualization tool VOSviewer to reveal the collaboration situation of global researchers. We set the minimum number of publications for an author to two. Based on this criteria, 38 authors met this threshold out of a total of 192 authors. 
Figure \ref{fig11} shows the chart with some unconnected nodes representing authors with a minimum of two publications but without collaboration with any of the other 37 authors.  
Figure \ref{fig12} shows in turn the chart with all nodes connected representing only authors with a collaboration.

In both Figure \ref{fig11} and Figure \ref{fig12}, each node represents an author. A link between two nodes represents a co-author relationship. To be specific, the bigger the node, the broader and more contributions made by the author. 
To obtain a better view of all the authors in our collaboration network, we analyze Figure \ref{fig11} containing all 38 authors. In Figure \ref{fig11}, there are various subgraphs with different colors, each subgraph represents a cluster and there are 11 clusters.  Authors with the same color show that they are in the same cluster and may have more cooperation with each other. Based on the size of nodes, the most prominent authors with a high number of collaborations and contributions include Richard Hawkins, Ibrahim Habli, Tim Kelly from the University of York, Barbara Gallina from Malardalen University, and Carmen Carlan from Fortiss GmbH research institute

Cluster 1 with the color red represents the research team from the University of Pennsylvania and shows that the majority of the publications in ACP have been done within the team without many collaborations with other authors outside of the University of Pennsylvania. Cluster 2 with the color blue presents the research team from Malardalen University, this cluster shows a connection between Barbara Gallina in Cluster 3 and Carlan in Cluster 4 (Fortiss GMBH) which represents a collaboration between different organizations. Cluster 3 with the colour green represents the research team from the University of York. This cluster represents the cluster having authors with the most collaboration with other authors from other universities. Cluster 5 shows the collaborative work that exists among Shuichiro Yamamoto (Nagoya University), Yutaka Matsuno in Japan, and Kenji Taguchi all affiliated with Japan. 
A noteworthy observation is cluster 6 which shows the collaboration that exists between both Ewen Denney and Ganesh Pai both affiliated with SGT / NASA Ames Research Center. This cluster shows no link or collaboration with other authors or affiliations. 

Based on this generated co-author chart that Figure \ref{fig11} depicts, the interconnections between nodes and clusters are a bit limited, which suggests that the majority of the authors in assurance case patterns might not be collaborating enough with other authors from other affiliations. 
Hence there is a need to encourage collaborative work to ensure the development of newer approaches and advancement in the field of assurance case patterns. 

\smallskip
\noindent\fbox{%
    \parbox{\linewidth}{%
    \smallskip
    In summary, for \textbf{RQ2}, The United Kingdom and the University of York rank first as the highest contributors in the field of assurance case patterns. The top institutions of authors contributing to the field of ACPs consisted of both academic institutions and research centres like Fortiss Gmbh and NASA. A key observation is that the United States of America has the highest number of diverse institutions in the author’s affiliation distribution. However, only one of its institutions emerged in the top five for author’s affiliation.
European countries emerged as the main contributors in the field of assurance case patterns. However, we noticed the absence of European countries such as Russia. We also noticed the absence of some countries from Asia, South America, and North America, and the total absence of countries from Africa and Oceania in our geographical landscape distribution. We attributed this absence to the language barrier and to the possibility that researchers with nationalities of these countries are affiliated with other institutions in foreign countries prominent in the field of ACPs. Also, we deduced that authors in the field of ACPs might not be collaborating enough with other authors from other affiliations.
Hence there is a need for active collaboration and cooperation among researchers from all countries or continents to speed up the development of newer approaches in the field of assurance case patterns especially in under-represented regions.

}}
\subsection{RQ3: {What are the potential future research contributions on assurance case pattern research?}}

\begin{figure*}[htbp]
    \centering
    \includegraphics[width=1.1\linewidth]{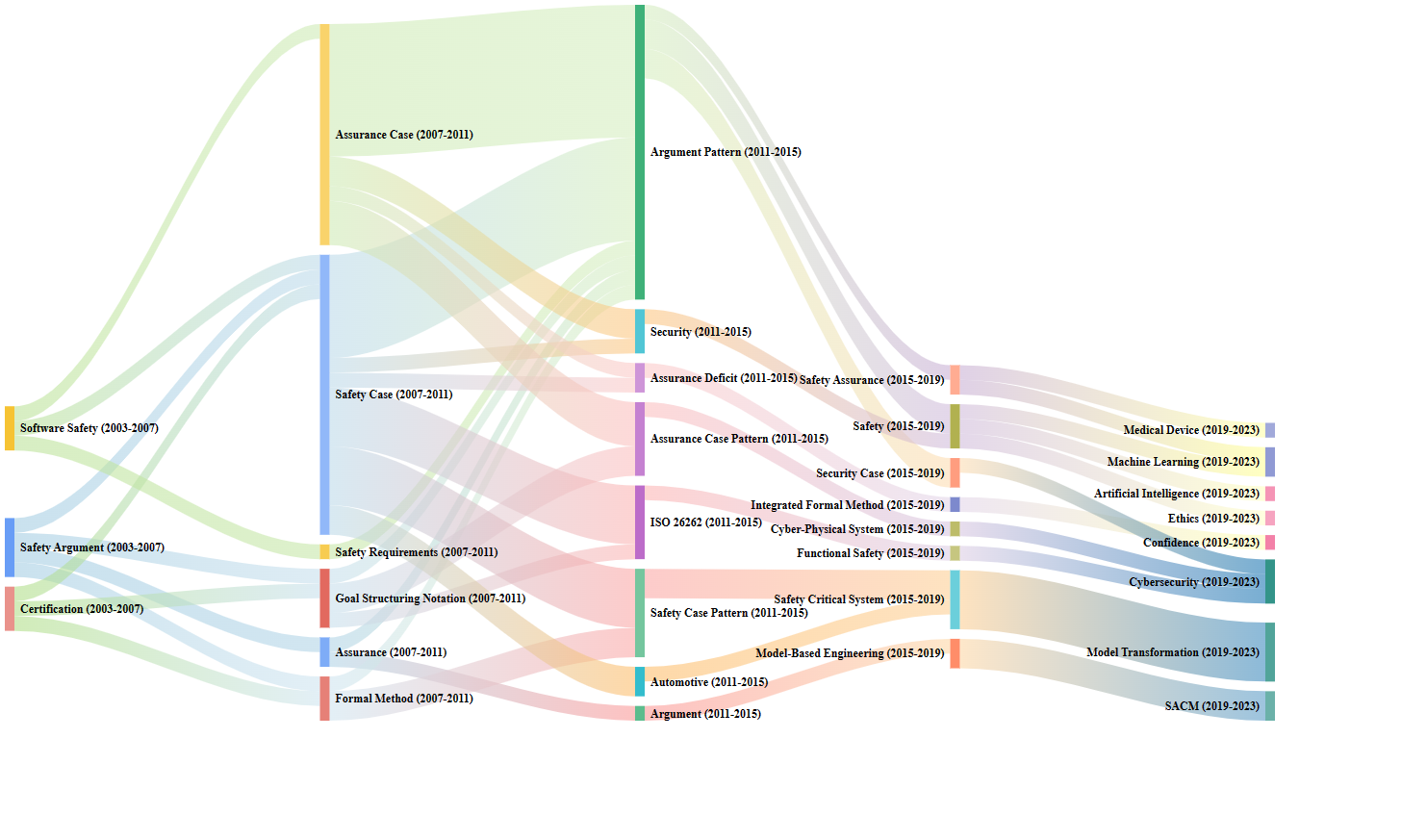}
    \caption{Evolution of research in assurance case pattern from January 2003 - October 2023}
    \label{fig13}
\end{figure*}

To explore the past, present, and potential future research directions on assurance case pattern research, we relied on a Sankey diagram that we created using Google Charts \cite{b48}. Figure \ref{fig13} depicts that diagram. The Sankey diagram shows the thematic evolution of the field of assurance case pattern within the last two decades. The blocks in the Sankey diagram represent the different paradigms and timelines in the field of ACP. As shown in the chart, researchers have explored diverse ACP-related themes over the years. %This suggests many directions for future works. 

In the Sankey diagram, the early years of 2003-2007 show the keywords \textbf{\textit{``software safety''}},\textbf{\textit{ ``safety argument''}} and \textbf{\textit{``certification''}} which suggests a focus on the use of safety arguments to support the certification of software systems. 
In the subsequent timeline of 2007-2011, the length of the block for \textbf{\textit{``assurance case''}} and \textbf{\textit{``safety case''}} signifies a wide adoption of assurance cases specifically safety cases to support assurance of system safety requirements Also, the keyword \textbf{\textit{``goal structuring notation''}} (GSN) shows the popularity of GSN as a notation for representing assurances cases which led to its standardization in 2011. In addition, this period shows the emergence of approaches for the formalization of assurance case patterns.  

Around 2011-2015, the chart shows that \textbf{\textit{``argument patterns''}}, \textbf{\textit{``safety case pattern''}}, \textbf{\textit{``assurance case patterns''}} have become mainstream approaches to support the reuse of argument structures to ease the generation of assurance cases. During this period argument patterns were also proposed to identify and mitigate assurance deficits in safety arguments. Furthermore, the keyword \textbf{\textit{``security''}} during this timeline suggests the emergence of ACPs that support security as another property of interest for system assurance. The introduction of the ISO 26262 standard in 2011 to provide functional safety guidelines in the automotive industry also coincides with the focus of ACP during this timeline to support and target the safety assurance of automotive vehicles. 

The period between 2015-2019 show the keywords \textbf{\textit{``safety critical system''}}, \textbf{\textit{``cyber-physical system''}}, \textbf{\textit{``security case''}}, \textbf{\textit{``functional safety''}}, \textbf{\textit{``safety assurance''}} and \textbf{\textit{``model-based engineering''}}. 
These keywords suggest a continuous focus on the safety assurance of safety-critical systems. It also presents the rise in both security and safety assurance of more complex systems such as cyber-physical systems developed using model-based engineering. The majority of the studies in assurance case patterns during this period proposed different automated approaches to extract arguments and evidence from design models of complex systems to instantiate ACP. 

Recently, under the current timeline of 2019-2023, the chart shows \textbf{\textit{``Medical Devices''}}, \textbf{\textit{``Machine Learning''}}, \textbf{\textit{``Artificial Intelligence''}}, \textbf{\textit{``Ethics''}},\textbf{\textit{``Confidence''}}, \textbf{\textit{``Cybersecurity''}}, \textbf{\textit{``Model Transformation''}}, and \textbf{\textit{``SACM''}} as the stand-out keywords. The wide application of artificial intelligence, specifically machine learning in various domains has brought about various systems having a machine learning component. Modern medical devices used in health care are ML-enabled and to this end, ACP is being proposed to provide the framework to argue assurance of a desired property and improve confidence in medical devices and other ML-enabled systems. 
\newline
Also, the ML components of most ML-enabled systems are usually black boxes without much knowledge about their decision-making process, this has presented an emerging requirement in ethical considerations of ML-enabled systems different from the safety and security assurance of systems.
Furthermore, the keyword \textbf{\textit{``cybersecurity''}} signifies a shift from security assurance to cybersecurity assurance and data privacy probably due to the rise of web-based systems. The terms \textbf{\textit{``model transformation''}} and \textbf{\textit{``SACM''}} signify the current approach of transforming information obtained from system design models as evidence and arguments to instantiate ACP to automatically create model-based AC.

Finally, from the keywords shown in the current timeline of 2019 - 2023, we can identify some of the potential future research directions on ACP research. 
%These directions may include the development of newer AC patterns to support the assurance of systems created with emerging AI technologies to increase confidence in their use and the development of patterns to support newer desired system properties as technology evolves.
We further discuss some of these directions in the remainder of this section.

\subsubsection{\textbf{Fine-tuning assurance case patterns}}
With the recent advancements of Large Language Models (LLM) that promote Generative AI, various domains have assumed a prominent role in leveraging LLM to tackle complex tasks. In the realm of software engineering, LLMs are currently being employed for a range of downstream tasks such as software defect prediction \cite{b135}, automated program repair \cite{b134}, and code generation \cite{b133}. In the field of safety assurance, we envision the application of LLMs to capture the internal structure of ACPs and the connections between their elements. This would support fine-tuning the argumentation structures that assurance case patterns capture in the current version of the system at hand to automatically generate assurance cases for the subsequent versions of that system. Hence, thanks to LLMS fine-tuning abilities, it may be possible in the future to manually create an assurance case of a version of a system and to automatically infer the assurance cases of all the subsequent versions of that system. This has the potential to progressively eliminate the manual creation and maintenance of assurance cases as systems evolve. 

\subsubsection{\textbf{Assurance case pattern to argue Fairness, Bias, and Explainability in ML-enabled systems}}
The widespread integration of artificial intelligence, especially machine learning, across diverse domains has led to the inclusion of machine learning components in many systems. Figure \ref{fig13} underscores the recent emphasis on assuring the performance of these systems, especially in the medical domain where machine learning models are used to diagnose patients and make decisions for medical treatment \cite{b3, b4}.
 However, owing to the intricate and black-box nature of machine learning components, ensuring confidence in the properties of machine learning-enabled systems is a non-trivial task. Recently, few studies \cite{b31, b78} have proposed the assurance of ethical considerations in machine learning-enabled systems. This suggests a growing need for the development of new assurance case patterns to support and provide sufficient confidence in the assurance of additional properties such as the absence of bias, fairness in decision-making and clarity in the decision-making process of ML-enabled systems.
\subsubsection{\textbf{Assurance case pattern with Prediction and Advice}}
Assurance deficits refer to "any knowledge gap that prohibits total or perfect confidence" \cite{b49} in an assurance case. Different studies \cite{b38, b40} in the literature have proposed the use of assurance case patterns to identify and mitigate these deficits. To measure the confidence in assurance cases created by ACPs, 
As future work, we foresee the development of a “\textit{trust parameter}” which would serve as a measure to predict and estimate the level of confidence in assurance cases generated from assurance case patterns.

\smallskip
\noindent\fbox{%
    \parbox{\linewidth}{%
    \smallskip
    In summary, for \textbf{RQ3}, We analyzed the evolution of the most influential keywords and potential future research directions in the field of assurance case pattern. Based on our findings, we envisage future research directions in the use of generative AI  to fine-tune ACPs to support incremental system assurance. We also envision the development of new ACPs to provide higher confidence in the assurance of emerging non-functional requirements such as ethics, fairness and bias.
}}

%\section{DISCUSSION}%

\section{THREATS TO VALIDITY}
\label{sec:6}
We adopt the classification of Zhou et al. \cite{b58} and Wohlin et al. \cite{b59} to discuss the threats to the validity of our work. 

\subsection{Internal Validity}

Hundreds of thousands of primary studies are usually surveyed and analyzed in most bibliometric analyses often due to the less stringent selection strategy employed. While this approach guarantees a high count of primary studies for the bibliometric analysis, it may introduce a higher likelihood of noise in the chosen studies. 

In our study, we have followed the guidelines proposed in \cite{b43,b44} to ensure a stringent selection strategy consisting of six phases to eliminate any occurrence of noisy data in our chosen studies. In addition, we have also employed the snowballing technique to identify additional relevant primary studies not found during the database-driven search across five scientific databases. Hence, our bibliometric analysis ensures completeness in the data search and selection process with minimal occurrence of noisy data. 

\subsection{Conclusion Validity}

In our study, the search for primary studies has been done from January 2003 to early October 2023. It is crucial to point out that the search for primary studies ended in early October 2023 which means that our search might miss a few studies published between October 2023 and December 2023. 

In addition, as explained above, we utilized the Connected Papers tool to perform snowballing. However, Connected Papers “\textit{could not find enough papers to create a graph }” for four papers comprised in the start set we used for snowballing. This was possibly due to the recency of these four papers. However, to mitigate this, we have followed all the systematic review and bibliometric analysis guidelines in \cite{b14,b43,b44} and ensured the transparency and reproducibility  of our work.

\section{CONCLUSION}
\label{sec:7}
In summary, this paper provides a bird’s eye view of the field of assurance case patterns using bibliometric analysis to help researchers and other stakeholders in the field of ACP to understand the evolutionary trend, and research direction and make informed decisions towards future research directions. In this regard, this paper presents a bibliometric analysis of 92 papers published in the last two decades (2003 - 2023) within the field of assurance case patterns. 
We used VosViewer \cite{b15}, Tableau \cite{b16}, Google Charts \cite{b48}, and Microsoft Excel for analysis and visualization of primary studies information from three perspectives. These perspectives are the trends defining the scientific literature, the diversity of scientific contributions, and the potential future research directions in this field. 

Results show that the number of publications from 2003 has varied with a peak in 2014. SAFECOMP emerges as the top choice venue for publications. The top three active countries are the United Kingdom, the United States of America, and Germany. The top three active institutions are the University of York (United Kingdom), Fortiss GMBH (Germany), and Malardalen University (Sweden). Besides, the hot topics that the primary studies cover include assurance cases, assurance case patterns, safety-critical systems, cyber-physical systems, safety, security, model-based engineering, machine learning, ethics, automotive, medical devices, and formal methods. 
\newline


\bibliographystyle{ACM-Reference-Format}
\bibliography{main}

\clearpage
\onecolumn
\appendix

\section{Appendices}
Table \ref{tab4} provides information regarding the 92 primary studies we included in our bibliometric analysis. That information consists of:  the authors of the study, its publication year, its title, its venue acronym, and the search method (e.g., snowball, database-driven) we used to find that study.
%Start the appendix with the ``\verb|appendix|'' command:

\subsection{List of Primary Studies}
\label{PryStudy}
{\footnotesize
\begin{longtable}[htbp]{p{0.7cm}|p{8.5cm}|p{2.5cm}|p{1.2cm}|p{1.7cm}}
\caption{List of Primary Studies}\\
\label{tab4}\\
\hline
 \textbf{No.} &
 \textbf{Study Title} & \textbf{Author(s) \& Publication Year} & \textbf{Venue} & \textbf{Search Method} \\
\hline
\endfirsthead

\multicolumn{5}{c}%
{{\bfseries \tablename\ \thetable{} -- continued from previous page}} \\
\hline
\textbf{No.} &
\textbf{Study Title} & \textbf{Author(s) \& Publication Year} & \textbf{Venue} & \textbf{Search Method} \\
\hline
\endhead

\hline \multicolumn{5}{|r|}{{Continued on next page}} \\ \hline
\endfoot

\hline \hline
\endlastfoot

1 & A Case Study on Safety Cases in the Automotive Domain: Modules, Patterns, and Models & Wagner et al. (\citeyear{b6}) & ISSRE & Database-Driven \\
\hline

2 & A design and implementation of an assurance case language & Matsuno Yutaka (\citeyear{b42})& DSN & Database-Driven \\
\hline

3 & A formal basis for safety case patterns & Denney \& Pai (\citeyear{b39}) & SAFECOMP & Database-Driven \\
\hline

4 & A framework to support generation and maintenance of an assurance case & Chung-Ling et al. (\citeyear{b54}) & ISSREW & Database-Driven \\
\hline

5 & A Generic Goal-Based Certification Argument for the Justification of Formal Analysis & Habli \& Kelly (\citeyear{b74}) & ENTCS & Snowballing \\
\hline

6 & A Layered Argument Strategy for Software Security Case Development & Biao et al. (\citeyear{b22}) & ISSREW & Database-Driven \\
\hline

7 & A method to generate reusable safety case argument-fragments from compositional safety analysis & Šljivo et al. (\citeyear{b75}) & JSS & Snowballing \\
\hline

8 & A New Approach to creating Clear Safety Arguments & Hawkins et al. (\citeyear{b49}) & SCSC & Snowballing \\
\hline

9 & A Pattern for Arguing the Assurance of Machine Learning in Medical Diagnosis Systems & Picardi et al. (\citeyear{b3}) & SAFECOMP & Database-Driven \\
\hline

10 & A Pattern to Argue the Compliance of System Safety Requirements Decomposition & Oliveira et al. (\citeyear{b76}) & SugarLoafPLoP & Snowballing \\
\hline

11 & A Pattern-Based Approach towards the Guided Reuse of Safety Mechanisms in the Automotive Domain & Khalil et al. (\citeyear{b77}) & IMBSA & Database-Driven \\
\hline

12 & A Pragmatic Approach to Reasoning about the Assurance of Safety Arguments & Weaver et al. (\citeyear{b63}) & SCS & Snowballing \\
\hline

13 & A principles-based ethics assurance argument pattern for AI and autonomous systems & Porter et al. (\citeyear{b78}) & Springer AI and Ethics & Database-Driven \\
\hline

14 & A safety case pattern for model-based development approach & Ayoub et al. (\citeyear{b57}) & NFM & Database-Driven \\
\hline

15 & A Safety Case Pattern for Systems with Machine Learning Components & Wozniak et al. (\citeyear{b79}) & SAFECOMP & Database-Driven \\
\hline

16 & A safety-case approach to the ethics of autonomous vehicles & Menon \& Alexander (\citeyear{b32}) & Safety and Reliability & Snowballing \\
\hline

17 & A Security Argument Pattern for Medical Device Assurance Cases & Finnegan \& McCaffery (\citeyear{b26}) & ISSREW & Database-Driven \\
\hline

18 & A Security Property Decomposition Argument Pattern for Structured Assurance Case Models & Jaskolka et al. (\citeyear{b80}) & EuroPLoP & Database-Driven \\
\hline

19 & A Systematic Approach for Developing Software Safety Arguments & Hawkins \& Kelly (\citeyear{b41}) & ISSC & Snowballing \\
\hline

20 & A Systematic Approach to Justifying Sufficient Confidence in Software Safety Arguments & Ayoub et al. (\citeyear{b64}) & SAFECOMP & Snowballing \\
\hline

21 & An Approach to Assure Dependability Through ArchiMate & Yamamoto Shuichiro (\citeyear{b81}) & SAFECOMP & Database-Driven \\
\hline

22 & An Assurance Case Pattern for the Interpretability of Machine Learning in Safety-Critical Systems & Ward \& Habli (\citeyear{b82}) & SAFECOMP & Database-Driven \\
\hline

23 & An evaluation of argument patterns based on data flow & Yamamoto Shuichiro (\citeyear{b83}) & ICT-EURASIA & Database-Driven \\
\hline

24 & An evaluation of argument patterns to reduce pitfalls of applying assurance case & Yamamoto \& Matsuno (\citeyear{b62}) & ASSURE & Database-Driven \\
\hline

25 & Applying Safety Case Pattern to Generate Assurance Cases for Safety-Critical Systems & Chung-Ling \& Wuwei (\citeyear{b56}) & HASE & Database-Driven \\
\hline

26 & Applying the Goal Structuring Notation (GSN) to Argue Compliance of Equipment with the European EMC Directive & Ghatge et al. (\citeyear{b84}) & IEEE Letters on Electromagnetic Compatibility Practice and Applications  & Database-Driven \\
\hline

27 & Arguing from Hazard Analysis in Safety Cases: A Modular Argument Pattern & Gleirscher \& Carlan (\citeyear{b85}) & HASE & Database-Driven \\
\hline

28 & Arguing on Software-Level Verification Techniques Appropriateness & Carlan et al. (\citeyear{b86}) & SAFECOMP & Snowballing \\
\hline

29 & Argument patterns for multi-concern assurance of connected automated driving systems & Warg \& Skoglund (\citeyear{b37}) & CERTS & Database-Driven \\
\hline

30 & Argumentation pattern: An approach to issuing software reliability case & Boxuan \& Minyan (\citeyear{b87}) & EITRT & Database-Driven \\
\hline

31 & Assurance argument patterns and processes for machine learning in safety-related systems & Picardi et al. (\citeyear{b88}) & SafeAI & Database-Driven \\
\hline

32 & Assurance Case Pattern using SACM Notation & Selviandro Nungki (\citeyear{b53}) & ICoICT & Database-Driven \\
\hline

33 & Assurance Case Patterns for Cyber-Physical Systems with Deep Neural Networks & Kaur et al. (\citeyear{b36}) & SAFECOMP & Database-Driven \\
\hline

34 & Assurance Cases for Block-Configurable Software & Hawkins et al. (\citeyear{b89}) & SAFECOMP & Snowballing \\
\hline

35 & Assurance Cases for Proofs as Evidence & Chaki et al. (\citeyear{b90}) & PCC & Snowballing \\
\hline

36 & Assurance of Automotive Safety - A Safety Case Approach & Palin \& Habli (\citeyear{b5}) & SAFECOMP & Database-Driven \\
\hline

37 & Assuring Safety for Component Based Software Engineering & Conmy \& Bate (\citeyear{b91}) & HASE & Database-Driven \\
\hline

38 & Automated Method for Assurance Case Construction from System Design Models & Hartsell et al. (\citeyear{b52}) & ICSRS & Database-Driven \\
\hline

39 & Automating Pattern Selection for Assurance Case Development forCyber-Physical Systems & Ramakrishna et al. (\citeyear{b92}) & SAFECOMP & Database-Driven \\
\hline

40 & Automotive safety case pattern & Macher et al. (\citeyear{b91}) & EuroPLoP & Database-Driven \\
\hline

41 & Combining GSN and STPA for Safety Arguments & Hirata \& Nadjm-Tehrani (\citeyear{b92}) & SAFECOMP & Snowballing \\
\hline

42 & Composition of Safety Argument Patterns & Denney \& Pai  (\citeyear{b93}) & SAFECOMP & Database-Driven \\
\hline

43 & Computer-Aided Generation of Assurance Cases & Wang et al. (\citeyear{b94}) & SAFECOMP & Database-Driven \\
\hline

44 & Confidence Arguments for Evidence of Performance in Machine Learning for Highly Automated Driving Functions & Burton et al. (\citeyear{b7}) & SAFECOMP & Database-Driven \\
\hline

45 & Constructing Security Cases Based on Formal Verification of Security Requirements in Alloy & Zeroual et al. (\citeyear{b95}) & SAFECOMP & Database-Driven \\
\hline

46 & CyberGSN: A Semi-formal Language for Specifying Safety Cases & Beyene \& Carlan (\citeyear{b96}) & DSN-W & Database-Driven \\
\hline

47 & Deriving Safety Case Fragments for Assessing MBASafe's Compliance with EN 50128 & Gallina et al. (\citeyear{b97}) & SPICE & Snowballing \\
\hline

48 & Design and implementation of GSN patterns: A step toward assurance case language & Matsuno Yutaka (\citeyear{b98}) & IPSJ & Database-Driven \\
\hline

49 & Developing Assurance Cases for D-MILS Systems & Hawkins et al. (\citeyear{b99}) & HiPEAC & Snowballing \\
\hline

50 & Enabling Cross-Domain Reuse of Tool Qualification Certification Artefacts & Gallina et al. (\citeyear{b100}) & SAFECOMP & Snowballing \\
\hline

51 & Enhancing state-of-the-art safety case patterns to support change impact analysis & Carlan \& Gallina (\citeyear{b8}) & ESREL & Database-Driven \\
\hline

52 & Enhancing the Cyber Resilience of Critical Infrastructures through an Evaluation Methodology Based on Assurance Cases & Koelemeijer Dorien (\citeyear{b101}) & Procedia Computer Science & Database-Driven \\
\hline

53 & Ethics in conversation: Building an ethics assurance case for autonomous AI-enabled voice agents in healthcare & Kaas et al. (\citeyear{b31}) & TAS & Database-Driven \\
\hline

54 & Evidence arguments for using formal methods in software certification & Denney \& Pai (\citeyear{b102}) & ISSREW & Database-Driven \\
\hline

55 & Evolution of Formal Model-Based Assurance Cases for Autonomous Robots & Gleirscher et al. (\citeyear{b103}) & SEFM & Database-Driven \\
\hline

56 & Experiences with Assurance Cases for Spacecraft Safing & Nguyen \& Ellis (\citeyear{b104}) & ISSRE & Database-Driven \\
\hline

57 & ExplicitCase: Tool-Support for Creating and Maintaining Assurance Arguments Integrated with System Models & Carlan et al. (\citeyear{b105}) & ISSREW & Database-Driven \\
\hline

58 & Facilitating construction of safety cases from formal models in Event-B & Prokhorova et al. (\citeyear{b106}) & Information and Software Technology  & Database-Driven \\
\hline

59 & General Development Framework and Its Application Method for Software Safety Case & Fuping et al. (\citeyear{b55}) & Journal of Software & Database-Driven \\
\hline

60 & Generation of Safety Case Argument-Fragments from Safety Contracts & Šljivo et al. (\citeyear{b65}) & SAFECOMP & Snowballing \\
\hline

61 & Hazard Contribution Modes of Machine Learning Components & Colin et al. (\citeyear{b107}) & AAAI & Snowballing \\
\hline

62 & Identifying and implementing security patterns for a dependable security case - From security patterns to D-case & Patu \& Yamamoto (\citeyear{b108}) & CSE & Database-Driven \\
\hline

63 & Integrated Formal Methods for Constructing Assurance Cases & Carlan et al. (\citeyear{b40}) & ISSREW & Snowballing \\
\hline

64 & Justifying the transition from trustworthiness to resiliency via generation of safety cases & Chung-Ling et al. (\citeyear{b109}) & SNPD & Database-Driven \\
\hline

65 & Justifying the validity of safety assessment models with safety case patterns & Sun et al. (\citeyear{b110}) & ISSC & Database-Driven \\
\hline

66 & Model Based System Assurance Using the Structured Assurance Case Metamodel & Wei et al. (\citeyear{b35}) & JSS & Snowballing \\
\hline

67 & Model-based Generation of Hazard-driven Arguments and Formal Verification Evidence for Assurance Cases & Yan et al. (\citeyear{b111}) & MODELSWARD & Snowballing \\
\hline

68 & Model-Connected Safety Cases & Retouniotis et al. (\citeyear{b112}) & IMBSA & Snowballing \\
\hline

69 & On using results of code-level bounded model checking in assurance cases & Carlan et al. (\citeyear{b113}) & SAFECOMP & Database-Driven \\
\hline

70 & Parameterised Argument Structure for GSN Patterns & Matsuno \& Taguchi (\citeyear{b66}) & ICQS & Database-Driven \\
\hline

71 & Patterns for Integrating NIST 800-53 Controls into Security Assurance Cases & Viger et al. (\citeyear{b38}) & SAFECOMP & Database-Driven \\
\hline

72 & Preventing recurrence of industrial control system accident using assurance case & Napolano et al. (\citeyear{b114}) & ISSREW & Database-Driven \\
\hline

73 & Product-line assurance cases from contract-based design & Nei et al. (\citeyear{b115}) & JSS & Database-Driven \\
\hline

74 & SACS - A pattern language for safe adaptive control software & Hauge et al. (\citeyear{b116}) & PLoP & Database-Driven \\
\hline

75 & Safe \& sec case patterns & Taguchi et al. (\citeyear{b117}) & SAFECOMP & Database-Driven \\
\hline

76 & Safety argument pattern language of safety-critical software & Wang et al. (\citeyear{b118}) & DSA & Database-Driven \\
\hline

77 & Software Reliability Case Development Method Based on the 4+1 Principles & Shihao et al. (\citeyear{b30}) & ICRMS & Database-Driven \\
\hline

78 & Software safety assurance - what is sufficient? & Hawkins \& Kelly (\citeyear{b119}) & ISSC & Snowballing \\
\hline

79 & Software Safety Certification Framework Based on Safety Case & Zeng et al. (\citeyear{b120}) & CSSS & Database-Driven \\
\hline

80 & Structural analysis of safety case arguments in a model-based development environment & Zechner \& Huhn (\citeyear{b121}) & MBEES & Database-Driven \\
\hline

81 & Support for safety case generation via model transformation & Chung-Ling et al. (\citeyear{b29}) & ACM SIGBED Review & Database-Driven \\
\hline

82 & The Need for a Weaving Model in Assurance Case Automation & Hawkins et al. (\citeyear{b122}) & AUJ & Snowballing \\
\hline

83 & Tool support for assurance case development & Denney \& Pai (\citeyear{b51}) & ASE & Database-Driven \\
\hline

84 & Tool-Supported Safety-Relevant Component Reuse: From Specification to Argumentation & Šljivo et al. (\citeyear{b123}) & AEiC & Snowballing \\
\hline

85 & Towards a case-based reasoning approach for safety assurance reuse & Ruiz et al. (\citeyear{b124}) & SAFECOMP & Database-Driven \\
\hline

86 & Towards assurance for plug \& Play medical systems & King et al. (\citeyear{b2}) & SAFECOMP & Database-Driven \\
\hline
87 & Towards Developing Safety Assurance Cases for Learning-Enabled Medical Cyber-Physical Systems & Bagheri et al. (\citeyear{b4}) & SafeAI & Snowballing \\
\hline

88 & Towards Goal-Based Software Safety Certification Based on Prescriptive Standards & Stensrud et al. (\citeyear{b125}) & WoSoCER & Database-Driven \\
\hline

89 & Towards safety case integration with hazard analysis for medical devices & Wardziski \& Jarzbowicz (\citeyear{b126}) & SAFECOMP & Database-Driven \\
\hline

90 & Uniform model interface for assurance case integration with system models & Wardziski \& Jones (\citeyear{b127}) & SAFECOMP & Database-Driven \\
\hline

91 & Using Process Models in System Assurance & Hawkins et al. (\citeyear{b128}) & SAFECOMP & Snowballing \\
\hline

92 & Weaving an Assurance Case from Design: A Model-Based Approach & Hawkins et al. (\citeyear{b50}) & HASE & Snowballing \\
\hline
\end{longtable}
}

\subsection{Publication Venue and Acronym}
\label{tab5}
{\footnotesize
\begin{longtable}[htbp]{p{0.3\textwidth}|p{0.6\textwidth}}
\caption{Publication Venue Names and Acronyms} \\
\hline
\textbf{Acronym} & \textbf{Publication Venue Name} \\
\hline
\endfirsthead
\multicolumn{2}{c}%
{\tablename\ \thetable\ -- \textit{Continued from previous page}} \\
\hline
\textbf{Acronym} & \textbf{Publication Venue Name} \\
\hline
\endhead
\hline \multicolumn{2}{r}{\textit{Continued on next page}} \\
\endfoot

AAAI & Association for the Advancement of Artificial Intelligence Conference \\
\hline

AEiC & Ada-Europe International Conference on Reliable Software Technologies \\
\hline

ASE & Automated Software Engineering \\
\hline

ASSURE & International Workshop on Assurance Cases for Software-Intensive Systems \\
\hline

AUJ & Ada User Journal \\
\hline

CERTS & International Workshop on Security and Dependability of Critical Embedded Real-Time Systems \\
\hline

CSE & International Conference on Computational Science and Engineering \\
\hline

CSSS & International Conference on Computer Science and Service System \\
\hline

DSA & International Conference on Dependable Systems and Their Applications \\
\hline

DSN & International Conference on Dependable Systems and Networks \\
\hline

DSN-W & International Conference on Dependable Systems and Networks Workshops \\
\hline

EITRT & International Conference on Electrical and Information Technologies for Rail Transportation \\
\hline

ENTCS & Electronic Notes in Theoretical Computer Science  \\
\hline

ESREL & European Conference on Safety and Reliability \\
\hline

EuroPLoP & European Conference on Pattern Languages of Programs \\
\hline

HASE & International Symposium on High-Assurance Systems Engineering \\
\hline

HiPEAC & International Conference on High Performance Embedded Architectures and Compilers \\
\hline

ICoICT & International Conference on Information and Communication Technology \\
\hline

ICQS & International Conference on Quality Software \\
\hline

ICRMS & International Conference on Reliability, Maintainability, and Safety \\
\hline

ICSRS & International Conference on System Reliability and Safety \\
\hline

ICT-EURASIA & EurAsian Conference on Information and Communication Technology \\
\hline

IMBSA & International Symposium on Model-Based Safety Assessment \\
\hline

IPSJ & Journal of Information Processing \\
\hline

ISSC & International Conference on System Safety  \\
\hline

ISSRE & International Symposium on Software Reliability Engineering \\
\hline

ISSREW & International Symposium on Software Reliability Engineering Workshops \\
\hline

JSS & Journal of Systems and Software \\
\hline

MBEES & Model-based development of embedded systems Workshop \\
\hline

MODELSWARD & International Conference on Model-Driven Engineering and Software Development \\
\hline

NFM & International Symposium on NASA Formal Methods \\
\hline

PCC & Proof Carrying Code Workshop \\
\hline

PLoP & Conference on Pattern Languages of Programs \\
\hline

SafeAI & The AAAI’s Workshop on Artificial Intelligence Safety \\
\hline

SAFECOMP & International Conference on Computer Safety, Reliability, and Security \\
\hline

SCS & Workshop on Safety critical systems and software \\
\hline

SSS & Safety critical system symposium \\
\hline

SEFM & International Conference on Software Engineering and Formal Methods \\
\hline

SNPD & International Conference on Software Engineering, Artificial Intelligence, Networking and Parallel/Distributed Computing \\
\hline

SPICE & International Conference on Software Process Improvement and Capability Determination \\
\hline

SugarLoafPLoP & The Latin American Conference on Pattern Languages of Programs \\
\hline

TAS & International Symposium on Trustworthy Autonomous Systems \\
\hline

WoSoCER & International Workshop on Software Certification \\
\hline
\endlastfoot

\end{longtable}
}
\end{document}